\documentclass[longauth]{aa}
\usepackage[utf8]{inputenc}
\usepackage[varg]{txfonts}
\usepackage[breaklinks, colorlinks, citecolor=blue, linkcolor=blue]{hyperref}
\usepackage[usenames, dvipsnames]{color}
\usepackage{amsmath}
\usepackage{graphicx}
\usepackage[english]{babel}
\usepackage{siunitx}
\usepackage{float}
\usepackage{url}
\usepackage{longtable}
\usepackage{fancyhdr}
\usepackage{natbib}
\usepackage{booktabs}
\usepackage[normalem]{ulem}
\usepackage[version=4]{mhchem}
\usepackage{gensymb}
\makeatletter
\def\instrefs#1{{\def\scsep{\def\scsep{,}}\@for\w:=#1\do{\scsep\ref{inst:\w}}}}
\renewcommand{\inst}[1]{\unskip$^{\instrefs{#1}}$}
\renewcommand{\autoref}
        {\def\equationautorefname{Eq.}%
         \def\figureautorefname{Fig.}%
         \def\sectionautorefname{Sect.}%
         \def\subsectionautorefname{Sect.}%
         \def\subsubsectionautorefname{Sect.}%
         \orgautoref}
\renewcommand*\aa@pageof{, page \thepage{} of \pageref*{LastPage}}
\makeatother

\newcommand{\host}{LTT\,3780}
\newcommand{\planetb}{LTT\,3780\,b}
\newcommand{\planetc}{LTT\,3780\,c}
 
\newcommand{\sdist}{$d_{\star}$}

\newcommand{\steff}{$T_\mathrm{eff}$}
\newcommand{\logsg}{$\log\,g_{\star}$}
\newcommand{\sfeh}{$\mathrm{[Fe/H]}$}

\newcommand{\sm}{$M_{\star}$}
\newcommand{\sr}{$R_{\star}$}

\newcommand{\svrotsini}{$\mathit{v}_\mathrm{rot} \sin\,i_{\star}$}

\newcommand{\sprot}{$P_\mathrm{rot}$}

\newcommand{\pvtcb}{\ensuremath{2458543.91281^{+0.00048}_{-0.00052}}}
\newcommand{\pvpb}{\ensuremath{0.768377^{+1.4e-06}_{-1.4e-06}}}
\newcommand{\pvkb}{\ensuremath{0.0324^{+0.0011}_{-0.001}}}
\newcommand{\pvbb}{\ensuremath{0.518^{+0.059}_{-0.089}}}
\newcommand{\pvasb}{\ensuremath{6.77^{+0.22}_{-0.21}}}
\newcommand{\pvab}{\ensuremath{0.01203^{+0.00054}_{-0.00053}}}
\newcommand{\pvtdurb}{\ensuremath{0.818^{+0.044}_{-0.022}}}
\newcommand{\pvincdegb}{\ensuremath{85.9^{+0.59}_{-0.50}}}
\newcommand{\pvmpb}{\ensuremath{2.34^{+0.24}_{-0.23}}}
\newcommand{\pvrpb}{\ensuremath{1.35^{+0.06}_{-0.06}}}
\newcommand{\pveb}{\ensuremath{0.064^{+0.075}_{-0.046}}}

\newcommand{\pvrhopb}{\ensuremath{5.24^{+0.94}_{-0.81}}}
\newcommand{\pvteqpb}{\ensuremath{1000^{+98}_{-100}}}
\newcommand{\pvgpb}{\ensuremath{12.6^{+1.8}_{-1.6}}}
\newcommand{\pvinspb}{\ensuremath{116^{+11}_{-10}}}
\newcommand{\pvtcc}{\ensuremath{2458546.8492^{+0.0016}_{-0.0017}}}
\newcommand{\pvpc}{\ensuremath{12.252131^{+7.2e-05}_{-6.4e-05}}}
\newcommand{\pvkc}{\ensuremath{0.0580^{+0.0016}_{-0.0016}}}
\newcommand{\pvbc}{\ensuremath{0.756^{+0.034}_{-0.031}}}
\newcommand{\pvasc}{\ensuremath{42.9^{+1.4}_{-1.3}}}
\newcommand{\pvac}{\ensuremath{0.0762^{+0.0034}_{-0.0034}}}
\newcommand{\pvtdurc}{\ensuremath{1.79^{+0.21}_{-0.21}}}
\newcommand{\pvincdegc}{\ensuremath{89.08^{+0.11}_{-0.13}}}
\newcommand{\pvmpc}{\ensuremath{6.29^{+0.63}_{-0.61}}}
\newcommand{\pvrpc}{\ensuremath{2.42^{+0.10}_{-0.10}}}
\newcommand{\pvec}{\ensuremath{0.115^{+0.07}_{-0.065}}}

\newcommand{\pvrhopc}{\ensuremath{2.45^{+0.44}_{-0.37}}}
\newcommand{\pvteqpc}{\ensuremath{397^{+39}_{-40}}}
\newcommand{\pvgpc}{\ensuremath{10.5^{+1.5}_{-1.3}}}
\newcommand{\pvinspc}{\ensuremath{2.88^{+0.28}_{-0.25}}}
\newcommand{\pvwdegb}{\ensuremath{251 \pm 66}}
\newcommand{\pvwdegc}{\ensuremath{250 \pm 36}}

\newcommand{\sdistva}[1][pc]{$22$\,#1}
\newcommand{\steffv}[1][$\mathrm{K}$]{${3360}\,{\pm}\,{51}$\,#1}
\newcommand{\logsgv}[1][$\mathrm{(cgs)}$]{${4.81}\,{\pm}\,{0.04}$\,#1}
\newcommand{\sfehv}[1][$\mathrm{dex}$]{${0.09}\,{\pm}\,{0.16}$\,#1}
\newcommand{\smv}[1][$\mathrm{M_{\odot}}$]{${0.379}\,{\pm}\,{0.016}$\,#1}
\newcommand{\srv}[1][$\mathrm{R_{\odot}}$]{${0.382}\,{\pm}\,{0.012}$\,#1}

\newcommand{\microm}{$\mathrm{\mu m}$}
\newcommand{\mps}{$\mathrm{m\,s^{-1}}$}

\newcommand{\kmps}{$\mathrm{km\,s^{-1}}$}
\newcommand{\cmpss}{$\mathrm{cm\,s^{-2}}$}

\newcommand{\ppb}{$b = (a/R_\star)\cos i_{\rm p}$}
\newcommand{\ppa}{$a_p$}
\newcommand{\ppi}{$i_p$}
\newcommand{\ppr}{$R_p$}
\newcommand{\ppm}{$M_p$}
\newcommand{\ppden}{$\rho_p$}
\newcommand{\ppgra}{$g_p$}
\newcommand{\ppteq}{$T_\mathrm{eq}$}
\newcommand{\ppins}{$S_p$}
\newcommand{\pbporb}{$P_\mathrm{b}$}
\newcommand{\pbr}{$R_\mathrm{b}$}
\newcommand{\pbm}{$M_\mathrm{b}$}
\newcommand{\pbden}{$\rho_\mathrm{b}$}
\newcommand{\pbins}{$S_\mathrm{b}$}
\newcommand{\pcporb}{$P_\mathrm{c}$}
\newcommand{\pcr}{$R_\mathrm{c}$}
\newcommand{\pcm}{$M_\mathrm{c}$}
\newcommand{\pcden}{$\rho_\mathrm{c}$}
\newcommand{\pcteq}{$T_\mathrm{eq}$}
\newcommand{\Mea}{$\mathrm{M_{\oplus}}$}
\newcommand{\Rea}{$\mathrm{R_{\oplus}}$}

\newcommand{\ppau}{au}
\newcommand{\ppdenu}{$\mathrm{g\,cm^{-3}}$}
\newcommand{\ppgrau}{$\mathrm{m\,s^{-2}}$}
\newcommand{\pptequ}{K}
\newcommand{\ppinsu}{$S_{\oplus}$}
\newcommand{\pbporbva}[1][d]{$0.77$\,#1}
\newcommand{\pbrv}[1][$\mathrm{R_{\oplus}}$]{\pvrpb\,#1}
\newcommand{\pbmv}[1][$\mathrm{M_{\oplus}}$]{\pvmpb\,#1}
\newcommand{\pbdenv}[1][${\rm g\,cm^{-3}}$]{\pvrhopb\,#1}
\newcommand{\pbteqv}[1][K]{\pvteqpb\,#1}
\newcommand{\pcporbva}[1][d]{$12.25$\,#1}
\newcommand{\pcrv}[1][$\mathrm{R_{\oplus}}$]{\pvrpc\,#1}
\newcommand{\pcmv}[1][$\mathrm{M_{\oplus}}$]{\pvmpc\,#1}
\newcommand{\pcdenv}[1][${\rm g\,cm^{-3}}$]{\pvrhopc\,#1}
\newcommand{\pcteqv}[1][K]{\pvteqpc\,#1}
\newcommand{\gaia}{\emph{\it Gaia}}
\newcommand{\corot}{\emph{\it CoRoT}}
\newcommand{\kepler}{\emph{{\it Kepler}}}
\newcommand{\ktwo}{\emph{{\it K2}}}
\newcommand{\tess}{\emph{{\it TESS}}}
\newcommand{\jwst}{\emph{{\it JWST}}}

\begin{document} 

\title{The CARMENES search for exoplanets around M dwarfs}
\subtitle{Two planets on opposite sides of the radius gap \\ transiting the nearby M dwarf \host}
\titlerunning{A planetary system around \host}
\author{G.~Nowak\inst{iac,ull}
\and R.~Luque\inst{iac,ull}
\and H.~Parviainen\inst{iac,ull}
\and E.~Pall\'{e}\inst{iac,ull}
\and K.~Molaverdikhani\inst{mpia}
\and V.\,J.\,S.~B\'{e}jar\inst{iac,ull}
\and J.~Lillo-Box\inst{cabesac}
\and C.~Rodr\'{i}guez-L\'{o}pez\inst{iaa}
\and J.\,A.~Caballero\inst{cabesac}
\and M.~Zechmeister\inst{iag}
% ...
}

\institute{
\label{inst:iac}Instituto de Astrof\'{i}sica de Canarias (IAC), 38205 La Laguna, Tenerife, Spain
\and \label{inst:ull}Departamento de Astrof\'{i}sica, Universidad de La Laguna (ULL), 38206, La Laguna, Tenerife, Spain
\and \label{inst:mpia}Max-Planck-Institut f\"{u}r Astronomie, K\"{o}nigstuhl 17, 69117 Heidelberg, Germany
\and \label{inst:cabesac}Centro de Astrobiolog\'{i}a (CSIC-INTA), ESAC, Camino bajo del castillo s/n, 28692 Villanueva de la Ca\~nada, Madrid, Spain
\and \label{inst:iaa}Instituto de Astrof\'{i}sica de Andaluc\'{i}a (IAA-CSIC), Glorieta de la Astronom\'{i}a s/n, 18008 Granada, Spain
\and \label{inst:iag}Institut f\"{u}r Astrophysik, Georg-August-Universit\"{a}t, Friedrich-Hund-Platz 1, 37077 G\"{o}ttingen, Germany
% ...
}

\date{Received 2 March 2020 / Accepted dd Month 2020}

\abstract{We present the discovery and characterisation of two transiting planets observed by the \emph{Transiting Exoplanet Survey Satellite} (\tess) orbiting the nearby (\sdist $\approx$ \sdistva), bright ($J \approx 9$\,mag) M3.5 dwarf \host{} (TOI--732). We confirm both planets and their association with \host{} via ground-based photometry and determine their masses using precise radial velocities measured with the CARMENES spectrograph. Precise stellar parameters determined from CARMENES high-resolution spectra confirm that \host{} is a mid-M dwarf with an effective temperature of \steff = \steffv, a surface gravity of \logsg = \logsgv, and an iron abundance of \sfeh = \sfehv, with an inferred mass of \sm = \smv{} and a radius of \sr = \srv. The ultra-short-period planet \planetb{} (\pbporb{} = \pbporbva{}) with a radius of \pbrv, a mass of \pbmv{}, and a bulk density of \pbdenv{} joins the population of Earth-size planets with rocky, terrestrial composition. The outer planet, \planetc{}, with an orbital period of \pcporbva{}, radius of \pcrv, mass of \pcmv{}, and mean density of \pcdenv{} belongs to the population of dense sub-Neptunes. With the two planets located on opposite sides of the radius gap, this planetary system is an excellent target for testing planetary formation, evolution, and atmospheric models. In particular, {\planetc} is an ideal object for atmospheric studies with the {\em James Webb Space Telescope}.}

\keywords{planetary systems -- techniques: photometric -- techniques: radial velocities -- stars: individual: LTT\,3470 -- stars: late-type}
\maketitle

\section{Introduction}
In a sequence of space-based transit surveys that commenced with the \corot{} mission \citep{2009A&A...506..411A} and continued with \kepler{} \citep{2010Sci...327..977B} and \ktwo{} \citep{2014PASP..126..398H}, the {\em Transiting Exoplanet Survey Satellite} \citep[\tess;][]{2015JATIS...1a4003R} is the first to cover nearly the entire sky in a search for transiting planets. 
This includes 2\,min short-cadence monitoring of almost all M dwarfs brighter than 15\,mag in the \tess{} bandpass. Based on \kepler{} results, \cite{2013ApJ...767...95D} showed that planets with radii below 4\,\Rea{} are almost the only type of planets orbiting M dwarfs. Planetary systems around bright \tess{} red dwarfs are therefore optimal laboratories for testing formation, evolution, and interior models of small planets (\ppr\,$\in$\,1--4 \Rea) that have no known counterparts in the Solar System. Detailed characterisation of small planets orbiting bright M dwarfs also allows optimal candidates to be selected for future in-depth atmospheric studies \citep[see e.g.][]{2013ApJ...764..182S,2018ApJ...856L..34B}.

As was shown by \cite{2017AJ....154..109F} and \cite{2018AJ....156..264F},  the radius distribution of small, close-in planets (with orbital periods $P <$ 100\,d) has a bi-modal structure with a gap around 1.7\,\Rea{} that separates the two main classes of small planets: presumably rocky super-Earths with radii centred at 1.2\,\Rea{}, and gas-dominated sub-Neptunes with radii centred at 2.4\,\Rea. For planets orbiting the same star, and hence exposed to the same irradiation, differences in the planetary bulk densities and atmospheric structures could then be mainly explained by differences in their masses and orbital distance. Planetary systems with two or more close-in, small planets located below and above the radius gap are therefore especially interesting for studying the formation, evolution, and atmospheric composition of small planets. So far only a few such systems have been characterised in terms of precise radii measured with space-based transit telescopes and masses determined via high-precision radial velocity (RV) measurements. These systems are: Kepler-10~bc \citep{2014ApJ...789..154D}, K2-106~bc \citep{2017AJ....153..271S,2017A&A...608A..93G}, HD~3167~bc \citep[K2-96~bc;][]{2017AJ....154..122C,2017AJ....154..123G}, GJ~9827~bcd \citep[K2-135~bcd;][]{2017AJ....154..266N,2018A&A...618A.116P}, K2-138~bcdef \citep{2018AJ....155...57C,2019A&A...631A..90L}, HD~15337~bc \citep[TOI-402~bc;][]{2019ApJ...876L..24G,2019A&A...627A..43D}, and K2-36~bc \citep{2019A&A...624A..38D}. 

Here, we present the discovery of two transiting \tess{} planets straddling the radius gap around the nearby, bright M dwarf \host{}. Given the brightness of the host star, these planets are suitable for detailed atmospheric characterisation with future ground- and space-based facilities. The paper is structured as follows: Section~\ref{sec-tess_photometry} presents the analysis of \tess{} photometry used for the discovery of planets around \host{}. In Section~\ref{ground_based_follow-up_observations}, ground-based observations of \host{} are presented, including seeing-limited transit photometry, high-resolution imaging, and high-resolution spectroscopy with CARMENES. Detailed analyses of the stellar properties of \host{} are presented in Section~\ref{sec-host}, while Section~\ref{sec-analysis_and_results} presents the joint analysis of all available data and the derived planetary properties. Finally, a discussion and conclusions are presented in Sections~\ref{sec-discussion} and ~\ref{sec-conclusions}.

\section{\emph{TESS} photometry}
\label{sec-tess_photometry}
\subsection{Space-based observations}
\label{subsec-space-based_observations}
\host{} (TOI--732, TIC~36724087) was observed by \tess{} in short cadence mode (2-min integrations) during cycle~1, sector~9 (camera~\#1, CCD~\#1) between 28 February and 25 March 2019, and it is expected to be observed again during the first year of the extended mission (cycle~3), sector~35 (camera~\#1) between 9 February and 7 March 2021.  In total, 21.736 days of science data were collected for \host{}. The 1.182-day gap in the \tess{} photometry between $\mathrm{BJD_{TDB}}$\,=\,2458555.54677 and $\mathrm{BJD_{TDB}}$\,=\,2458556.72869 was caused by the data download during perigee passage. Data collected between $\mathrm{BJD_{TDB}}$\,=\,2458543.22185 and $\mathrm{BJD_{TDB}}$\,=\,2458544.43991 (1.218\,day) and between $\mathrm{BJD_{TDB}}$\,=\,2458556.72869 and $\mathrm{BJD_{TDB}}$\,=\,2458557.85228 (1.123\,day) were excluded from the light curves of all targets on Camera~\#1 CCD~\#1 because of the strong scattered light at the beginning of orbits 25 and 26 that was problematic for systematic error correction and the subsequent transiting planet search.

\subsection{Transit search}
\label{subsec-transit_search}
\begin{figure*}
    \centering
    \includegraphics[width=\linewidth]{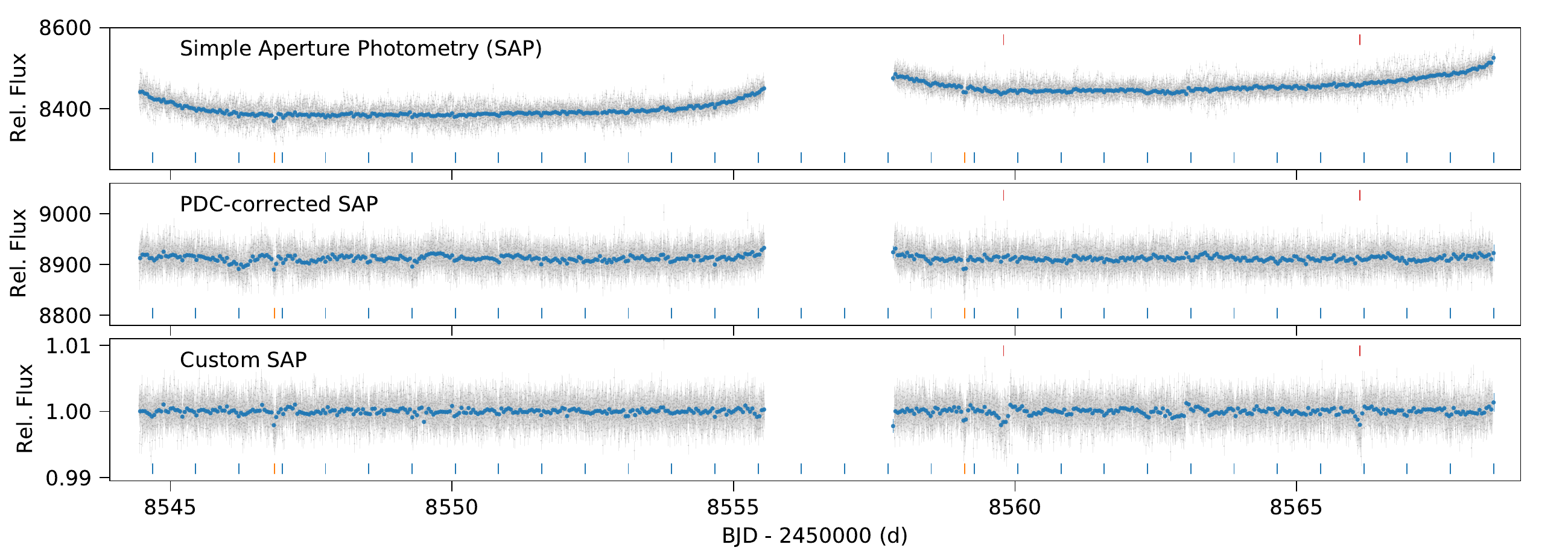}
    \caption{{\tess} data of {\host}.
    {\it Top panel:} Simple-aperture photometry from SPOC pipeline.
    {\it Middle panel:} PDC-corrected photometry from SPOC pipeline.
    {\it Bottom panel:} Custom-aperture photometry as in \citet{Hidalgo2020}. 
    Blue and orange ticks below the light curve mark the transits of the candidates TOI--732.01 (blue) and TOI--732.02 (orange). Red ticks above the light curves mark two dips that might correspond to single-transit events.\label{tess-dhs_lc}}
\end{figure*}

We downloaded from the Mikulski Archive for Space Telescopes\footnote{\url{https://mast.stsci.edu/portal/Mashup/Clients/Mast/Portal.html}} (MAST) the corresponding \tess{} light curve of \host{} produced by the Science Processing Operations Center \citep[SPOC;][]{SPOC} at the NASA Ames Research Center. For this target, SPOC provided simple aperture photometry \citep[SAP;][]{twicken:PA2010SPIE,2017ksci.rept....6M} and systematics-corrected photometry, a procedure consisting of an adaptation of the {\it Kepler} Pre-search Data Conditioning algorithm \citep[PDC;][]{Smith2012PASP..124.1000S, Stumpe2012PASP..124..985S, Stumpe2014PASP..126..100S} to \tess. The SPOC light curves generated by both methods are shown in the first two panels of Fig.~\ref{tess-dhs_lc}.

Additionally, we retrieved the \tess{} target pixel file (TPF) from MAST and performed a custom selection of pixels to build the optimal aperture that maximises the transit signals in the light curve (see Section~\ref{subsec-limits_on_photometric_contamination}). Following the methods described in \citet{Hidalgo2020}, we used our own analysis pipeline based on the {\tt everest}\footnote{\url{https://github.com/rodluger/everest}} pipeline \citep{Luger2016,Luger2018}, which applies the pixel level decorrelation (PLD) technique \citep{Deming2015} to extract from the raw light curve a final, flattened version (see bottom panel in Fig.~\ref{tess-dhs_lc}). To detect possible transit events, we used the box-fitting least squares (BLS) algorithm \citep{2002A&A...391..369K} on the flattened light curve to search for periodic signals. Once a signal was detected, we modelled the transit with \texttt{batman} \citep{Kreidberg2015} and removed it from the light curve. We iteratively used this procedure until no further signals were detected. We found two periodic signals at $0.7685 \pm 0.0007$ and $12.254 \pm 0.007$ days, respectively. These signals are associated with {\em TESS} objects of interest (TOIs) 732.01 and 732.02. Alerts were issued for the TOIs based on SPOC Data Validation reports \citep{2010ApJ...713L..87J,Twicken:DVdiagnostics2018,Li:DVmodelFit2019}, which identified the two transit signatures.

\addtocounter{footnote}{+1}
\subsection{Limits on photometric contamination}
\label{subsec-limits_on_photometric_contamination}
\begin{figure}
    \centering
    \includegraphics[width=\linewidth]{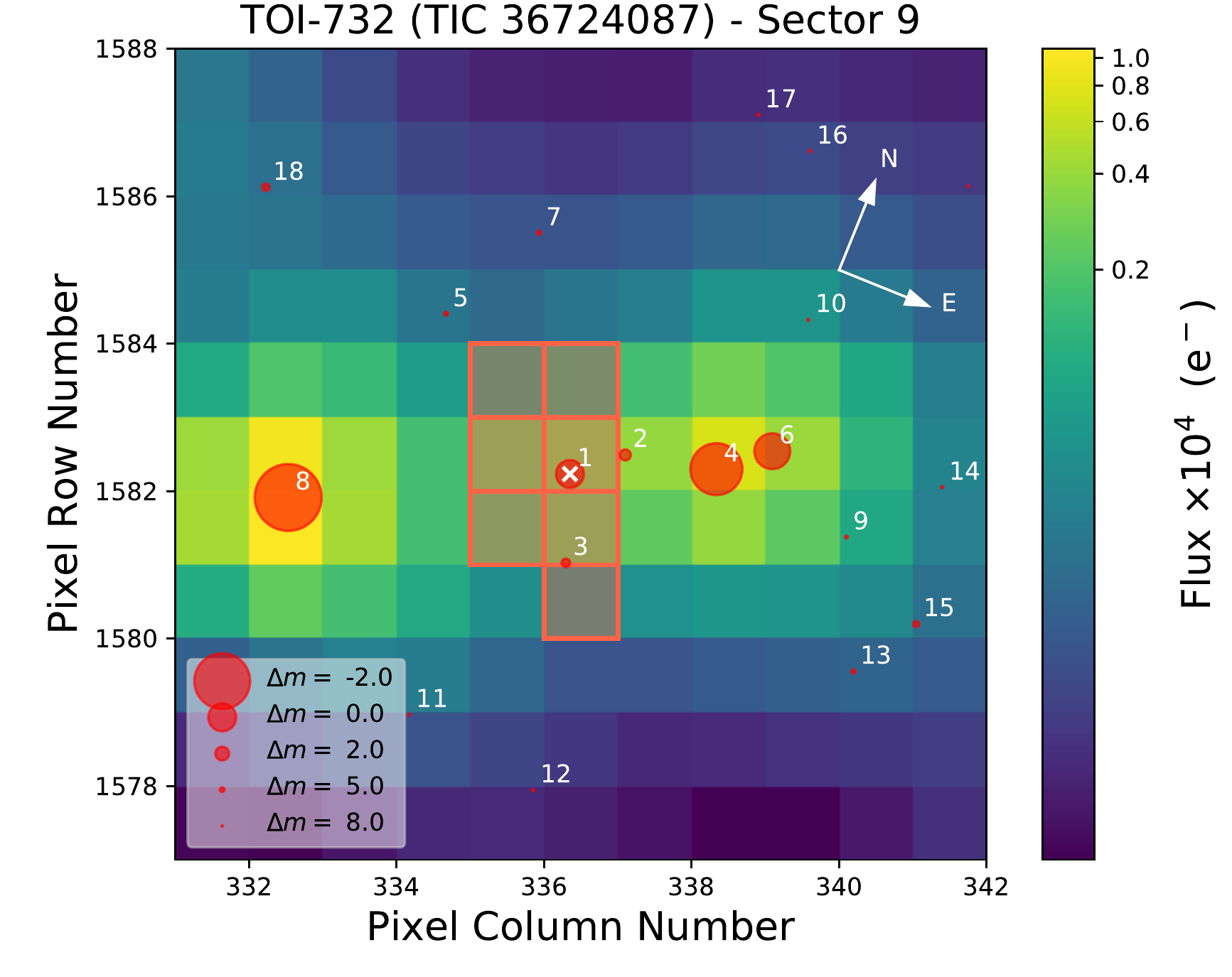}
    \caption{\tess{} image of {\host} in Sector~9 (created with \texttt{tpfplotter}\protect\footnotemark, \citealt{aller20}). The electron counts are colour-coded. The red bordered pixels are used in SAP. The size of the red circles indicates the \tess{} magnitudes of all nearby stars and \host{} (label \#1 with the "$\times$"). Positions are corrected for proper motions between \gaia{} DR2 epoch (2015.5) and \tess{} Sector~9 epoch (2019.2). The \tess{} pixel scale is 21\arcsec approximately.
    \label{figure-pht-tess-image}}
\end{figure}

The common proper motion companion to \host{}, namely \object{LP~729--55} (TIC~36724086, label~\#2 in Fig.~\ref{figure-pht-tess-image}\footnotetext{\url{https://github.com/jlillo/tpfplotter}}; see Section~\ref{subsec-host-binary}), was located just outside the aperture mask used to extract the light curve of \host{}. However, another close-in star, TIC~36724077 (\gaia{} DR2 3767281536635597568, 2MASS~J10183398--1143258), separated by 23.42\arcsec{} from \host{} and 3.4\,mag fainter than \host{} in the broad \gaia{} $G$ band (label \#3 in Fig.~\ref{figure-pht-tess-image}) was located within the aperture mask. Taking advantage of the similar spectral coverage of the \gaia{} $G_{R_{P}}$ band (630--1050\,nm) and \tess{} band (600--1000\,nm), we estimated the dilution factor for \tess{}, $D_{TESS} = 1/(1+F_{C}/F_{T})$, based on integrated \gaia{} $R_{P}$ mean fluxes of the contaminant star, TIC~36724077 ($F_{C} = 10476.9 \pm 17.5$), and \host{} itself ($F_{T} = 437973 \pm 662$), to be $D_{TESS} = 0.9766$. This means that TIC~36724077 dilutes the transit depths in the light curve of \host{} and hence decreases the apparent planet--star radius ratios. Therefore, to measure the unaffected radii of planets we fitted for the dilution factor in \tess{} photometry in the combined transit and RV analysis (see Section~\ref{subsec-analysis-joint_model}).
\addtocounter{footnote}{-1}

\subsection{A third single-transit planet?}
\label{subsec-single_transit_signal}
While we find no statistically significant signals for additional transits, once the two planets are removed, the light curves present two dips that might correspond to single-transit events, namely at 2458559.796 and 2458566.124 BJD. The presence and shape of these potential transits vary depending on the use of MAST flattened light curves or our custom flattening procedure, and in both cases the transit shape is not clear, so we cannot claim the presence of a third transiting planet in the system. The analysis of future {\em TESS} re-observations of this object should help solve this ambiguity. 

\section{Ground-based follow-up observations}
\label{ground_based_follow-up_observations}
\subsection{Seeing-limited transit photometry}
We acquired ground-based time-series follow-up photometry of \host{} as part of the {\em TESS} Follow-up Observing Program (TFOP)\footnote{\url{https://tess.mit.edu/followup/}} to attempt to ($i$) rule out nearby eclipsing binaries as potential sources of the {\em TESS} detection, ($ii$) detect the transit-like event on target to confirm the event depth and thus the {\em TESS} photometric deblending factor, ($iii$) refine the {\em TESS} ephemeris, ($iv$) provide additional epochs of transit centre time measurements to supplement the transit timing variation analysis, and ($v$) place constraints on transit depth differences across optical filter bands. We used the {\tt TESS Transit Finder}, which is a customised version of the {\tt Tapir} software package \citep{Jensen:2013}, to schedule our transit observations. Unless otherwise noted, the photometric data were extracted using the {\tt AstroImageJ} software package \citep{Collins:2017}. A summary of the photometric ground-based observations is shown in Table~\ref{tab:obs}\footnote{The LCOGT, TRAPPIST-North, SNO-T150, OAA, and MuSCAT2 light curves can be provided upon request to the first author.}.

\begin{table}[t]
    \caption{Observing log of ground-based photometric observations.}
    \centering
    \begin{tabular}{@{}lcccc@{}}
    \hline\hline
    \noalign{\smallskip}
    Telescope & Planet & Date  &  Filter & Coverage  \\
    \noalign{\smallskip}
    \hline
    \noalign{\smallskip}
TCS                 & .01 & 2019-12-28 & $g$ $r$ $i$ $z_s$  & 100 \% \\
                    & .01 & 2020-01-24 & $g$ $r$ $i$ $z_s$  & 100 \% \\
                    & .01 & 2020-01-27 & $g$ $r$ $i$ $z_s$  & 100 \% \\
                    & .01 & 2020-01-30 & $g$ $r$ $i$ $z_s$  & 100 \% \\
                    & .02 & 2019-12-10 & $g$ $r$ $i$ $z_s$  & 80 \% \\
                    & .02 & 2020-01-28 & $g$ $r$ $i$ $z_s$  & 100 \% \\
SNO-T150            & .02 & 2019-12-10 & $V$ $R$            & 100 \% \\
TRAPPIST-North      & .02 & 2019-12-10 & $z$                & 100 \% \\
LCO-CTIO            & .01 & 2019-06-09 & $z_s$              & 100 \% \\
                    & .01 & 2019-06-16 & $z_s$              & 100 \% \\
LCO-SAAO            & .01 & 2019-06-17 & $g_p$ $z_s$        & 100 \% \\
LCO-SSO             & .02 & 2020-01-04 & $B$                & 100 \% \\
OAA                 & .01 & 2020-02-01 & $Ic$               & 100 \% \\
    \noalign{\smallskip}
    \hline
    \end{tabular}
    \label{tab:obs}
\end{table}

\subsubsection{Las Cumbres Observatory network}
In total, four transits of the \host{} system where observed with the SINISTRO CCDs operating in the 1m telescopes of the Las Cumbres Observatory (LCOGT) network\footnote{\url{https://lco.global/}} \citep{Brown:2013}. For \planetb, two transits were observed from the Cerro Tololo Inter-American Observatory (CTIO) site using the $z_s$ filter on 9 and 16 June 2019. Exposure times were set to 45\,s and 70\,s, respectively, and an optimum aperture of 15.0\,pix (5.83\,\arcsec). A third transit was observed using the $g_p$ and $z_s$ filters from the South Africa Astronomical Observatory (SAAO) site on 17 June 2019 with exposure times of 140~s and the optimum apertures of 12.0\,pix (4.67\,\arcsec) and 15.0\,pix (5.83\,\arcsec) for filters $g_p$ and $z_s$, respectively. For \planetc, one transit observation was performed from Siding Spring Observatory (SSO) on 4 January 2020, using the $B$ filter, an exposure time of 140~s and an optimum aperture of 7\,pix (2.66\,\arcsec).

\subsubsection{TRAPPIST-North}
TRAPPIST-North at Oukaimeden Observatory in Morocco is a 60\,cm Ritchey-Chr\'{e}tien telescope, which has a thermo-electrically cooled 2k$\times$2k Andor iKon-L BEX2DD CCD camera with a field of view of 20\arcmin$\times$20\arcmin and pixel scale of 0.60\,\arcsec\,pix$^{-1}$ \citep{jehin2011,barkaoui2019}. We carried out a full-transit observation of TOI--732.02 on 10 December 2019 using a $z$ filter with an exposure time of 10\,s. We took 507 images and performed aperture photometry with an optimum aperture of 11~pixels (6.6\arcsec) and a point spread function (PSF) full width half maximum (FWHM) of 3.7\arcsec. We confirmed the event on the target star with a depth of $\sim$3.2\,ppt (parts per thousand) and occurring about 1\,h sooner with respect to the predicted ingress, but still within the expected uncertainty from an ephemeris derived from the {\em TESS} data only. We cleared all the stars of eclipsing binaries within the 2.5\arcmin around the target star.

\subsubsection{Sierra Nevada Observatory-T150 multicolour photometry}
T150 at Sierra Nevada Observatory (SNO) in Granada (Spain) is a 150\,cm Ritchey-Chr\'{e}tien telescope equipped with another thermo-electrically cooled 2k$\times$2k Andor iKon-L BEX2DD CCD camera with a field of view of 7.9\arcmin$\times$7.9\arcmin and pixel scale of 0.23\,\arcsec\,pix$^{-1}$. We carried out a full-transit observation of TOI--732.02 on 10 December 2019 with $R$ and $V$ filters (placed on a filter wheel) with an exposure time of 20\,s and 60\,s, respectively. We took 125 images (2$\times$2 binning) in both filters, and performed aperture photometry with an optimum aperture of 10\,pix (4.6\arcsec) in $R$ and 11\,pix (5.0\arcsec) in $V$ and a PSF FWHM of 2.3\,\arcsec and 2.7\arcsec, respectively. We confirmed the event on the target star in both filters, with similar depth of $\sim$3.2\,ppt and also about 1\,h earlier than the predicted ingress.

\subsubsection{Observatori Astronòmic Albanyà}
Additional photometric observations of TOI--732.01 were acquired on 1 February 2020 with the 0.4\,m telescope at the Observatori Astronòmic Albanyà (OAA), in Catalonia, Spain. The host star was observed for 391.7\,min with a Cousins $I_c$ filter and a Moravian G4-9000 camera with a field of view of 3056(H) $\times$ 3056(V) pixels covering 36.8\,\arcmin. We performed aperture photometry with an optimum aperture of 14\,pix (20.3\arcsec) and a PSF FWHM of 13.0\,\arcsec. We confirmed the event on the target with depth of $\sim$2.9\,ppt and occurring about 15 minutes later than the predicted ingress (within the expected uncertainty of the ephemeris derived from the TESS data only).

\subsubsection{Telescopio Carlos S\'anchez/MuSCAT2 multicolour photometry}
\label{section.muscat2}
We observed four transits of TOI--732.01 and two transits of TOI--732.02 with the MuSCAT2 multi-colour imager \citep{MuSCAT2} installed at the Telescopio Carlos S\'anchez (TCS), located at the Teide Observatory in Tenerife, Spain. The instrument carries out high-precision, simultaneous photometry in four colours ($g$, $r$, $i$, $z_\mathrm{s}$) with a pixel scale of 0.44\arcsec\,pix$^{-1}$. Observations were reduced with a dedicated MuSCAT2 pipeline \citep[see][for details]{Parviainen2019A&A...630A..89P}.

The exposure times varied from passband to passband and night to night depending on observing conditions and the observer's judgement. The shortest exposure times were of the order of 5\,s, and the longest of 60\,s. The aperture photometry was performed with an the optimum apertures of 5--6\arcsec, depending on the seeing and possible defocusing at a given observing night. The reduction included an initial detrending, after which all the light curves were down-sampled to a 1\,min cadence. The covariates used in the detrending were also down-sampled and stored, and used in the linear baseline model in the final joint light curve and RV modelling.

\subsection{Other light curves from public databases}
\label{subsec:phot}
We compiled photometric series obtained by long-time baseline automated surveys as in \mbox{\citet{DiezAlonso2019A&A...621A.126D}}. We were only able to retrieve data from the All-Sky Automated Survey for Supernovae \citep[ASAS-SN;][]{Kochanek2017PASP..129j4502K}, but not from other public catalogues, such as the All-Sky Automated Survey \citep[ASAS;][]{Pojmanski2002AcA....52..397P}, Northern Sky Variability Survey \citep[NSVS;][]{Wozniak2004AJ....127.2436W}, The MEarth Project \citep{MEarth}, the Catalina surveys \citep{Drake2014ApJS..213....9D}, or the Hungarian Automated Telescope Network \citep{Bakos2004PASP..116..266B}. The ASAS-SN dataset comprises 220 observations spanning about 2000\,d in $V$ band.

Additionally, \host{} is a candidate of the Super-Wide Angle Search for Planets  \citep[SuperWASP;][]{Pollacco2006PASP..118.1407P}. SuperWASP acquired more than 40\,000 photometric observations using a broad-band optical filter spanning two consecutive seasons from January to June 2013, and January to June 2014. For our search to detect long-term photometric modulations associated with the stellar rotation, we binned the data to one-day intervals, resulting in 191 epochs.

\subsection{High-contrast imaging}
\begin{figure*}
    \centering
    \includegraphics[width=\linewidth]{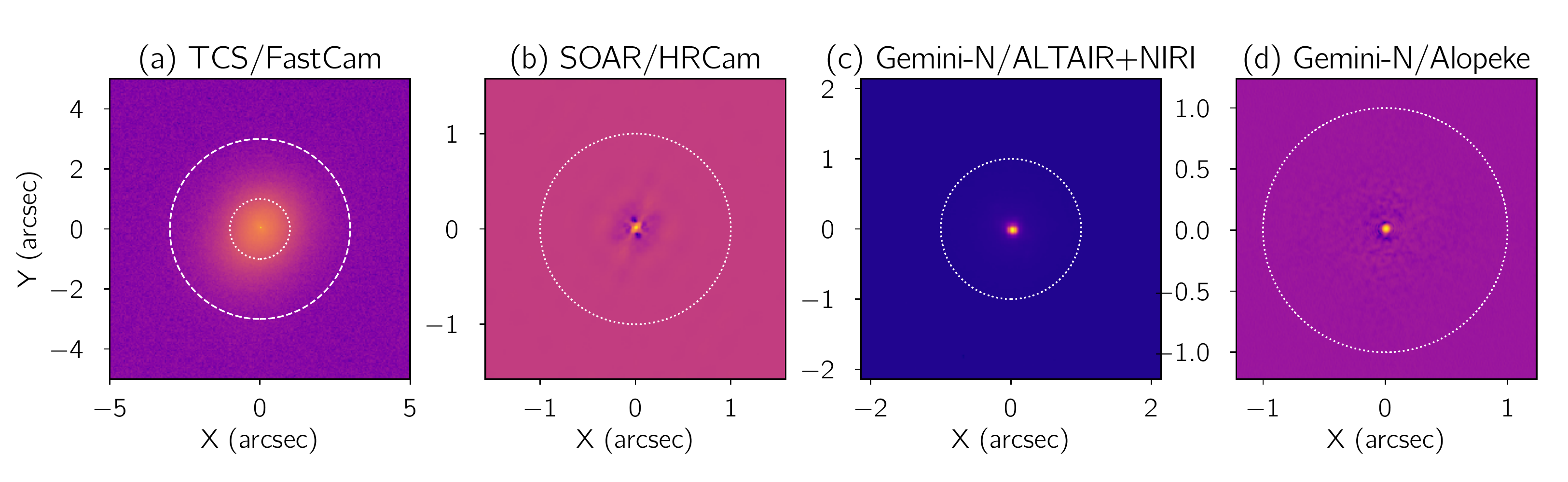}
    \caption{High-spatial-resolution images of \host{} from TCS/FastCam lucky imaging ({\it panel a}), SOAR/HRCam speckle imaging ({\it panel b}), Gemini North/ALTAIR+NIRI ({\it panel c}), and Gemini North/`Alopeke at 832\,nm ({\it panel d}). The dotted circle corresponds to 1\arcsec and the dashed circle to 3\arcsec separation. North is up and east is left.
    \label{figure-hci-soar_hrcam}}
\end{figure*}

\begin{figure}
    \centering
    \includegraphics[width=1\linewidth]{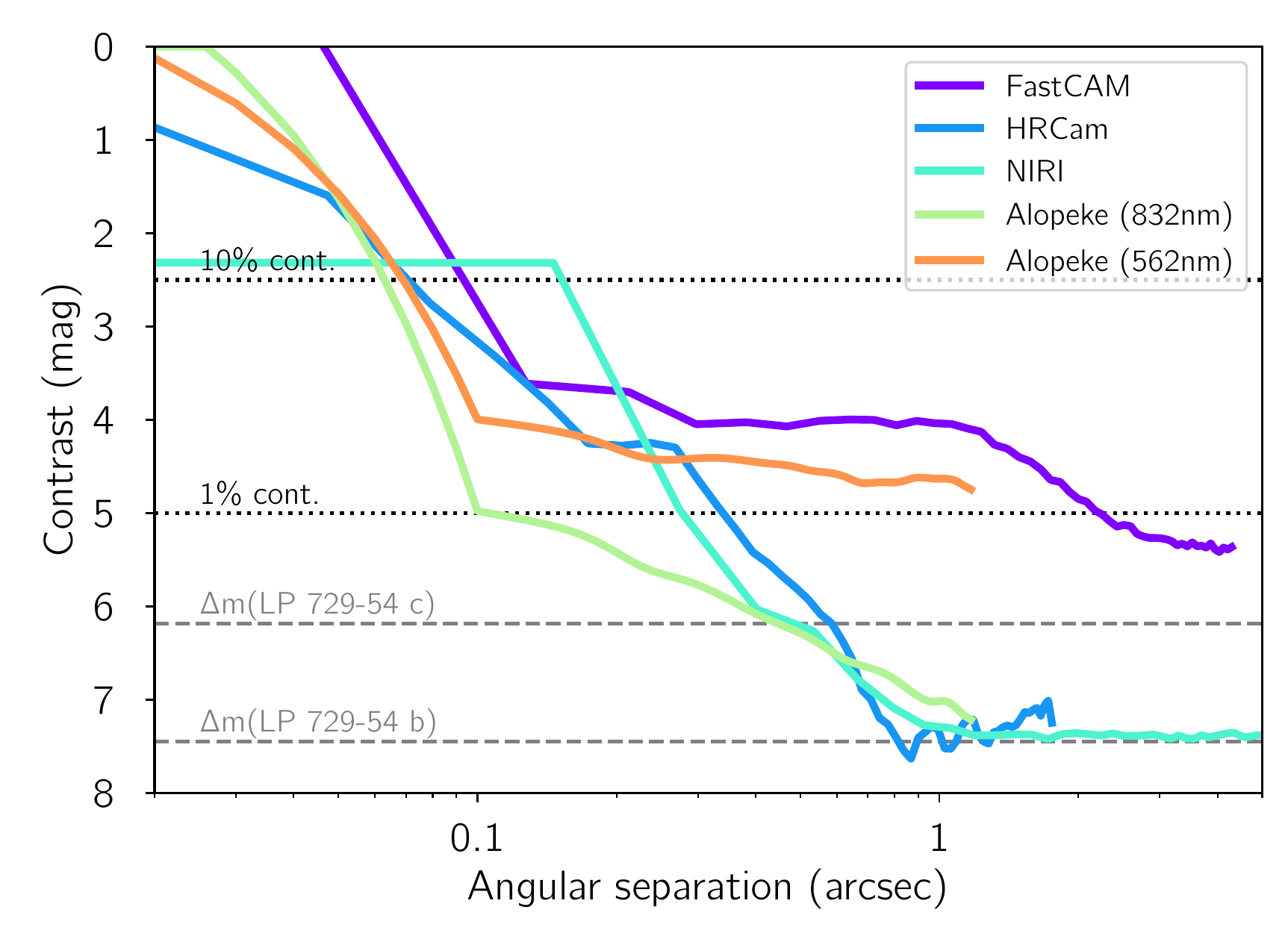}
    \caption{Sensitivity curves (5$\sigma$ limits) for all five high-spatial resolution images used in this work. The 1\,\% and 10\,\% contamination levels are marked as black dotted horizontal lines and the maximum magnitude contrast that a blended binary could have to mimic the transit depth of the two planets in the system are marked as grey dashed horizontal lines.
    \label{figure-sensitivity}}
\end{figure}

The presence of an unknown star within the same \tess{} pixel as the target can result in under-estimated planetary radii, caused by the additional light diluting the transit depth. These unaccounted stars could also potentially be the source of astrophysical false positives, although this was found to be unlikely for multi-planet transiting systems \citep{2012ApJ...750..112L}. To search for close-in companion stars (bound or unbound to the target star), and to estimate the potential contamination factor from such sources, we used high-contrast images of \host{} acquired with four instruments: 
(1) FastCam \citep{2008SPIE.7014E..47O} mounted on the 1.5\,m TCS (see Section~\ref{section.muscat2}); 
(2) HRCam \citep{2018PASP..130c5002T} installed on the 4.1\,m Southern Astrophysical Research (SOAR) telescope at CTIO; 
(3) `Alopeke\footnote{\url{https://www.gemini.edu/sciops/instruments/alopeke-zorro/}}, a high-resolution speckle interferometry instrument on the 8\,m Frederick C.~Gillett Gemini North telescope at Gemini North Observatory, Hawai'i, USA; 
and (4) Near InfraRed Imager and spectrograph \citep[NIRI,][]{2003PASP..115.1388H} coupled with the adaptive optics (AO) system facility, ALTAIR, mounted also on Gemini North.

\subsubsection{TCS/FastCam lucky imaging}
\host{} was observed  with the FastCam instrument on 22 May 2014 in the $I$ band as part of our high-resolution imaging campaign to identify and characterise the binary content of the CARMENES sample of M dwarfs \citep{2017A&A...597A..47C}. FastCam is a lucky imaging camera mounted on TCS and is equipped with a high readout speed and sub-electron noise L3CCD Andor 512$\times$512 detector, with a pixel size of 0.0425\,\arcsec, which provides a field of view of 21.2\arcsec $\times$ 21.2\arcsec. Ten blocks of 1\,000 individual frames of 50\,ms exposure time were obtained for this target. Data were processed using a dedicated pipeline developed by the Universidad Polit\'ecnica de  Cartagena group \citep[see][]{2010SPIE.7735E..0XL,2013MNRAS.429..859J}, which includes bias correction, the alignment and combination of images, and the selection of the best-quality images using the pixel position and value of the brightest speckle. Figure~\ref{figure-hci-soar_hrcam} (panel \textit{a}) shows the FastCam image resulting from selecting the 50\% best-quality images of the first 4\,000 frames. The corresponding 5$\sigma$ detection sensitivity curve is shown in Fig.~\ref{figure-sensitivity}. No additional source is detected with $\delta\,I<$~4\,mag down to the resolution limit of the telescope ($\sim$\,0.15\,\arcsec).

\subsubsection{SOAR/HRCam speckle imaging}
\host{} was observed with SOAR/HRCam speckle imaging on 12 December 2019 (UT) in $I$ band, a similar visible bandpass to that of \tess, to search for nearby sources. Further details of observations are available in \cite{2020AJ....159...19Z}. The region within 3\,\arcsec{} of \host{} was found to be devoid of nearby stars (see Fig.~\ref{figure-hci-soar_hrcam}, panel \textit{b}) within the 5$\sigma$ detection sensitivity of the observation, which is shown in  Fig.~\ref{figure-sensitivity}.

\subsubsection{Gemini North/`Alopeke speckle imaging}
\host{} was observed on 17 February 2020 (UT) at Gemini North with `Alopeke, the high-resolution speckle interferometry instrument. The star was observed simultaneously in two bandpasses centred at 562\,nm and 832\,nm (the latter is shown in panel \textit{d} of Fig.~\ref{figure-hci-soar_hrcam}). The resulting contrast curves are shown in Fig.~\ref{figure-sensitivity}, from which we deduced that \host{} does not have any close companion. At the distance of \host{}, our inner working angle is 0.37\,au at 562\,nm and 0.62\,au at 832\,nm, respectively, and our field of view ($r$ = 1.25\arcsec) extends out to 28\,au from the star. Given that \host{} is an M3.5\,V star, our contrast curves eliminate any companion object down to the M--L boundary in luminosity.

\subsubsection{Gemini-North/NIRI+ALTAIR AO imaging}
On 25 November 2019 (UT) we acquired AO images of \host{} with the Gemini North/NIRI+ALTAIR using the Br$\gamma$ filter (Gemini-North ID G0218) centred at 2.17\,\microm. We collected nine images, each with an integration time of 2.2\,s, and dithered the telescope between each exposure. Images were reduced following standard procedures, that is, correction for bad pixels, flat-fielding, subtraction of a sky background constructed from the dithered images, alignment of the star between frames, and co-addition of data. The Gemini North/NIRI+ALTAIR AO image of \host{} (see panel \textit{c} of Fig.~\ref{figure-hci-soar_hrcam}) revealed no close-in companions and the star appeared single to the limit of our resolution. Fig.~\ref{figure-sensitivity} presents the 5$\sigma$ contrast curve as a function of the angular separation from \host. We calculated the sensitivity to faint companions as a function of radius by injecting synthetic point spread functions at a range of magnitudes into the data, and measuring the significance at which they could be recovered. The data were of high quality, and we were sensitive to companions 5.0\,mag fainter than the target at just 270\,mas, and 7.4\,mag fainter than the target at separations greater than $\sim$1\arcsec.

\subsection{High-resolution spectroscopy}
\label{sec-hrs}
\subsubsection{3.5\,m Calar Alto/CARMENES}
\label{sec:observations.carmenes}
We obtained 52 high-resolution spectra of \host{} between 27 December 2019 (UT) and 19 February 2020 (UT) with the CARMENES instrument \citep{CARMENES,CARMENES18} mounted on the 3.5\,m telescope at the Calar Alto Observatory, Almer\'ia, Spain, as part of the guaranteed time observation program to search for exoplanets around M dwarfs \citep{Reiners17}. The CARMENES spectrograph has two channels, the visible (VIS) one covering the spectral range $0.52$--$\SI{0.96}{\micro\metre}$ and a near-infrared (NIR) channel covering the spectral range $0.96$--$\SI{1.71}{\micro\metre}$. 

Relative radial-velocity values, chromatic index (CRX), differential line width (dLW), and H$\alpha$ index values were obtained using {\tt serval}\footnote{\url{https://github.com/mzechmeister/serval}} \citep{SERVAL}. For each spectrum, we also computed the cross-correlation function (CCF) and its FWHM, contrast (CTR) and bisector velocity span (BVS) values, following \citet{2020A&A...636A..36L}. The RV measurements were corrected for barycentric motion, secular acceleration, and nightly zero-points. For more details, see \citet{Trifonov18} and \citet{Kaminski18}.

\subsubsection{Subaru/IRD}
We observed \host{} with the InfraRed Doppler instrument \citep[IRD,][]{2018SPIE10702E..11K} behind an AO system \citep[AO188,][]{2010SPIE.7736E..0NH} on the Subaru 8.2\,m telescope on Mauna Kea Observatories, as part of the Subaru IRD-{\tess} intensive follow-up project (S19A--069I). We took four spectra of \host{} on 10 December 2019 (UT) and one spectrum on 13 December 2019 simultaneously with laser-frequency comb spectra. Exposure times were 480\,s and a signal-to-noise ratio (S/N) at $\SI{1.0}{\micro\metre}$ was $\sim$100 for the four spectra, but the S/N was only $\sim $15 for the last one owing to thick clouds. We reduced the raw IRD frames of \host{} using the echelle package of {\tt iraf} for flat-fielding, scattered-light subtraction, aperture tracing, and wavelength calibration with the Th-Ar lamp spectra. For RV measurements requiring a more precise wavelength calibration, the wavelength was re-calibrated based on the emission lines of the combined laser frequency comb, which was injected simultaneously into both stellar and reference fibres during instrument calibrations. We injected these reduced spectra into the RV analysis pipeline for Subaru/IRD \citep{2020ApJ...890L..27H} and attempted to reproduce the intrinsic stellar template spectrum from all the observed spectra with instrumental profile deconvolution and telluric removal. Radial velocities were measured with respect to that template by forward-modelling of the observed individual spectral segments (each spanning 0.7--1.0\,nm). We found that the RV precision for the first-night spectra was typically 2.4\,m\,s$^{-1}$, while that of the second night was $\approx$19\,m\,s$^{-1}$ due to the low quality of the spectrum. We therefore discarded the latter measurement from the final analysis.

\subsubsection{NASA
Infrared Telescope Facility/iSHELL}
\label{sec:observations.ishell}
We obtained 77 five-minute spectra during seven nights for LP\,729-54 spanning 28\,days from January to February 2020 with the iSHELL spectrometer on the {NASA Infrared Telescope Facility} \citep[IRTF,][]{2016SPIE.9908E..84R}. Five-minute exposures were repeated 8-15 times within a night to reach a cumulative photon signal-to-noise ratio per spectral pixel at about $2.2\,\mu$m (at the approximate centre of the blaze for the middle order) varying from 152 to 205 to achieve a per-night RV precision of 6--10\,m\,s$^{-1}$. Spectra were reduced and RVs extracted using the methods outlined in \citet{2019AJ....158..170C}.

Due to the limited barycentre sampling of the iSHELL observations over the small observing window, the underlying stellar spectrum could not be well-isolated from other spectral features (namely tellurics). Therefore, a synthetic stellar spectrum was used to compute the RVs instead of deriving a more robust stellar template from the observations themselves. The overall RV scatter is consequently larger than expected given the S/N and RV information content of the observations. Two outliers (first and fourth night) were disregarded in the analysis, which can be recovered in the future with additional observations at different barycentre velocities.

The radial velocities collected with 3.5\,m Calar Alto/CARMENES, Subaru/IRD and IRTF/iSHELL instruments and their uncertainties are listed in Table~\ref{table-rvs}.

\section{Stellar parameters}
\label{sec-host}
\begin{table}
\centering
\small
\caption{Stellar parameters of \host.\label{tab:star}}
\begin{tabular}{@{}lcr@{}}
\hline\hline
\noalign{\smallskip}
Parameter                               & Value                 & Reference \\ 
\hline
\noalign{\smallskip}
\multicolumn{3}{c}{Name and identifiers}\\
\noalign{\smallskip}
Name                            & \object{LTT\,3470}    & \citet{1957cnes.book.....L}\\
                                & G~162--44             & \citet{Giclas71}\\
Karmn                           & J10185--117           & \citet{Carmencita}\\
TOI                             & 732                   & \tess{} Alerts\\
TIC                             & 36724087              & {\citet{2018AJ....156..102S}}\\
\noalign{\smallskip}
\multicolumn{3}{c}{Coordinates and spectral type}\\
\noalign{\smallskip}
$\alpha$ (J2015.5)              & 10:18:34.77       & {\it Gaia} DR2\\
$\delta$ (J2015.5)              & --11:43:04.1      & {\it Gaia} DR2\\
Sp. type                        & M3.5\,V           & {\citet{2003AJ....126.3007R}}\\
\noalign{\smallskip}
\multicolumn{3}{c}{Magnitudes}\\
\noalign{\smallskip}z§
$B$ [mag]                               & $14.68\pm0.04$        & UCAC4       \\
$g$ [mag]                               & $13.84\pm0.05$        & UCAC4       \\
$G_{BP}$ [mag]                          & $13.352\pm0.004$    & {\it Gaia} DR2       \\
$V$ [mag]                               & $13.14\pm0.04$        & UCAC4       \\
$r$ [mag]                               & $12.55\pm0.05$        & UCAC4       \\
$G$ [mag]                               & $11.8465\pm0.0005$    & {\it Gaia} DR2       \\
$i$ [mag]                               & $11.09\pm0.08$        & UCAC4       \\
$G_{RP}$ [mag]                          & $10.6583\pm0.0016$    & {\it Gaia} DR2       \\
$J$ [mag]                               & $9.01\pm0.03$         & 2MASS       \\
$H$ [mag]                               & $8.44\pm0.06$         & 2MASS       \\
$K_s$ [mag]                             & $8.20\pm0.02$         & 2MASS       \\
$W1$ [mag]                              & $8.04\pm0.02$         & AllWISE       \\
$W2$ [mag]                              & $7.880\pm0.019$         & AllWISE       \\
$W3$ [mag]                              & $7.771\pm0.019$         & AllWISE       \\
$W4$ [mag]                              & $7.58\pm0.17$         & AllWISE       \\
\noalign{\smallskip}
\multicolumn{3}{c}{Parallax and kinematics}\\
\noalign{\smallskip}
$\pi$ [mas]                             & $45.46\pm0.08$    & {\it Gaia} DR2             \\
$d$ [pc]                                & $22.00\pm0.04$    & {\it Gaia} DR2             \\
$\mu_{\alpha}\cos\delta$ [$\mathrm{mas\,yr^{-1}}$]  & $-341.411\pm0.11$ & {\it Gaia} DR2          \\
$\mu_{\delta}$ [$\mathrm{mas\,yr^{-1}}$]            & $-247.87\pm0.10$ & {\it Gaia} DR2          \\
$V_r$ [$\mathrm{km\,s^{-1}}]$           & $-0.44\pm0.09$    & {\citet{2018A&A...614A..76J}}    \\
$U$ [$\mathrm{km\,s^{-1}}]$             & $-14.89\pm0.06$   & This work      \\
$V$ [$\mathrm{km\,s^{-1}}]$             & $-22.00\pm0.08$   & {This work}      \\
$W$ [$\mathrm{km\,s^{-1}}]$             & $-35.08\pm0.07$   & {This work}      \\
Galactic population                     & Thin disc         & This work \\
\noalign{\smallskip}
\multicolumn{3}{c}{Photospheric parameters}\\
\noalign{\smallskip}
$T_{\mathrm{eff}}$ [K]                      & $3360\pm51$       & {This work}   \\
$\log g$                                    & $4.81\pm0.04$     & {This work}   \\
{[Fe/H]}                                    & $+0.09\pm0.16$    & {This work}   \\
$v \sin i_\star$ [$\mathrm{km\,s^{-1}}$]    & $<3.0$            & {\citet{2018A&A...614A..76J}}             \\
\noalign{\smallskip}
\multicolumn{3}{c}{Physical parameters}\\
\noalign{\smallskip}
$L$ [$10^{-4}\,L_\odot$]                & $167 \pm 3$       & {This work}       \\
$R$ [$R_{\odot}$]                       & $0.382 \pm 0.012$     & {This work}       \\
$M$ [$M_{\odot}$]                       & $0.379 \pm 0.016$     & {This work}       \\
$\rho$ [$\mathrm{g\,cm^{-3}}$]          & $9.6 \pm 1.0$         & {This work}\\
\noalign{\smallskip}
\hline
\end{tabular}
\tablebib{
    2MASS: \citet{2MASS};
    AllWISE: \citet{AllWISE};
    {\it Gaia} DR2: \citet{2018A&A...616A...1G};
    UCAC4: \citet{UCAC4}.
}
\end{table}

\subsection{The stellar host and its binary system}
\label{subsec-host-binary}
\host{} is a relatively bright ($J \, \approx \, 9.0$\,mag) M3.5\,V star at approximately 22\,pc \citep{2018A&A...616A...1G}. According to the Washington Double Star catalogue \citep{WDS}, it is the primary of the poorly investigated wide system LDS~3977 \citep{Luyten63}. The secondary, located at angular separation $\rho$ = 15.81$\pm$0.15\arcsec\ and position angle $\theta$ = 96.9$\pm$0.2\,deg (at epoch J2015.5), is \object{LP~729--55}, which is about 2.1\,mag fainter in the $J$ band and shares within $1 \sigma$ the same parallax and proper motions as \host{} \citep{2018A&A...616A...1G}. At the system heliocentric distance, $\rho$ translates into a projected physical separation of $s$ = 348$\pm$3\,au. \cite{2003AJ....126.3007R} assigned the spectral type M3.5\,V to the secondary from low-resolution spectroscopy. However, we consider that they meant the primary instead, whose spectral type agrees within 0.5\,dex with that derived by \citet{2005A&A...442..211S}, as well as with its effective temperature (see below). 
From accurate absolute magnitudes, colours, and luminosity and using various magnitude-, colour-, and luminosity-spectral type relationships available in the literature for solar-metallicity M-type stars, we estimate an m5.0$\pm$0.5\,V spectral type for the common proper companion secondary LP~729--55.

\subsection{Photospheric and physical parameters}
\label{subsec-host-parameters}
The photospheric parameters of \host{} were determined following \citet{Passegger19} using improved PHOENIX-ACES \citep{PHOENIX-ACES} stellar atmosphere models, which include a new equation of state to especially account for spectral features of low-temperature stellar atmospheres, as well as new atomic and molecular line lists. Effective temperature, surface gravity, and metallicity were derived assuming $v \sin i_\star = 2\,\mathrm{km\,s^{-1}}$ and a stellar age of 5\,Gyr \citep[see][]{Passegger19}. The latter two values are consistent with the rotational velocity upper limit determined by \citet{2018A&A...614A..76J} and an approximate solar age from the kinematic membership in the Galactic thin disc, using the same Galactocentric space velocity computation as \citet{CC16}.

To compute the physical parameters we followed the multi-step approach of \citet{2019A&A...625A..68S}. First, we determined the luminosity $L$ by integrating the photometric stellar energy distribution collected for the CARMENES targets \citep{Carmencita} with the Virtual Observatory Spectral energy distribution Analyser \citep{Bayo08} using parallactic distances from the \gaia{} DR2 catalogue \citep{2018A&A...616A...1G}. We then derived the radius $R$ and mass $M$ using the Stefan–Boltzmann's law and the empirical $M$-$R$ relation presented in \citet{2019A&A...625A..68S}, respectively.

We derived an effective temperature of $T_{\rm eff}=3360\pm51\,\mathrm{K}$, a stellar mass of $M_\star = 0.379 \pm 0.016\,\mathrm{M_{\odot}}$, and a radius of $R_\star = 0.382 \pm 0.012\,\mathrm{R_{\odot}}$, resulting in a stellar density of $\rho = 9.6\pm1.0\,\mathrm{g\,cm^{-3}}$. All derived values and additional stellar parameters can be found in Table~\ref{tab:star}.

\subsection{Stellar activity and rotation period}
\label{subsec-host-prot}
\begin{figure}
    \centering
    \includegraphics[width=\linewidth]{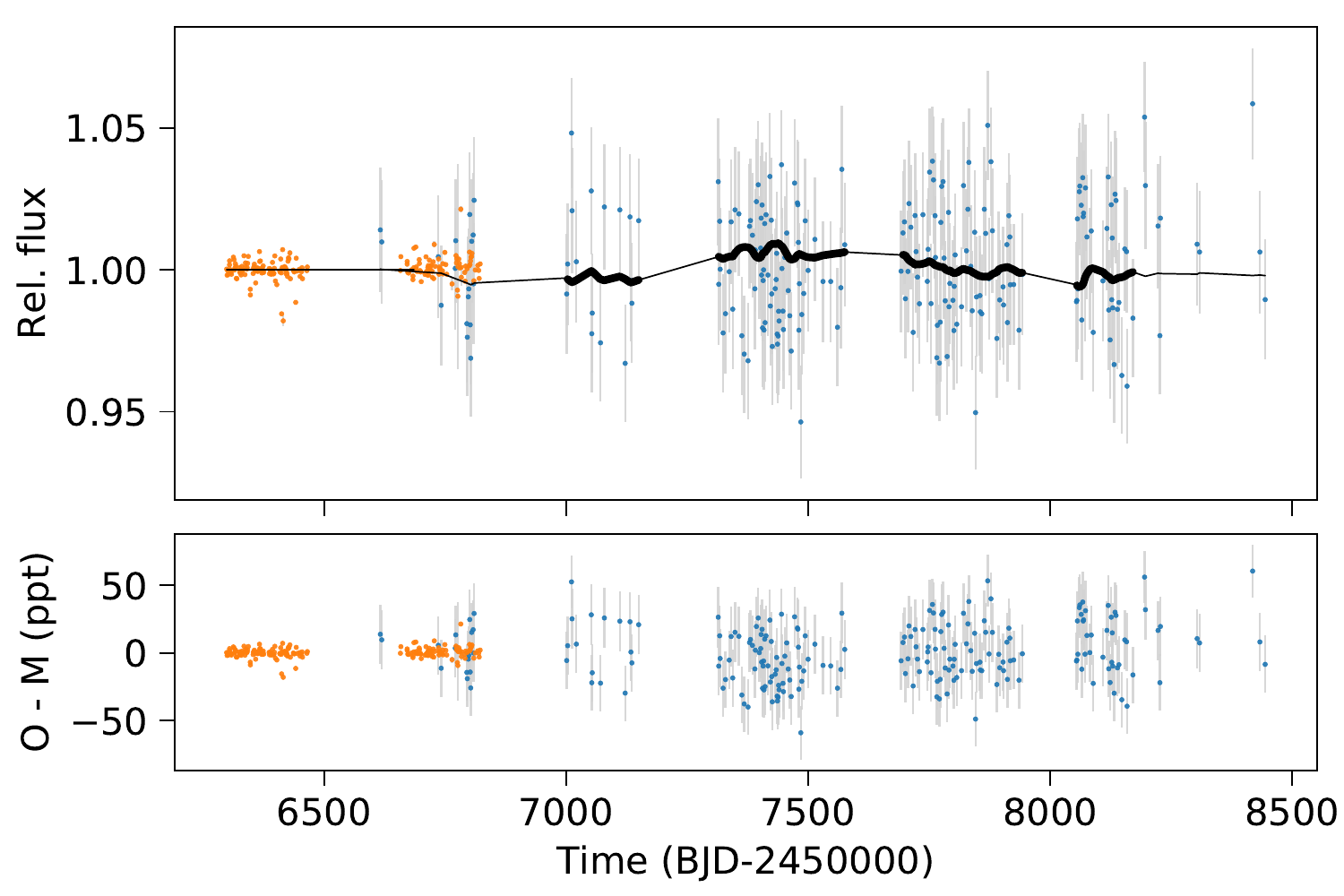}
    \caption{SuperWASP (orange) and ASAS-SN (blue) long-term photometric monitoring modelled with a quasi-periodic GP kernel defined as in \citet{2017AJ....154..220F}.
    \label{fig:asas_lc}}
\end{figure}

Using the stellar radius determined in Section~\ref{subsec-host-parameters} and presented in Table~\ref{tab:star} (\sr\,=\,\srv) and the upper limit for the stellar projected rotation velocity found by \citet[][\svrotsini\,<\,3\,\kmps]{2018A&A...614A..76J}, we calculated the stellar rotation period {\sprot} to be longer than 6\,d, and shorter than 100\,d for \svrotsini\,>\,0.2\,\kmps. We performed a search for the rotation period in the existing photometric data of \host{} from the SuperWASP and ASAS-SN surveys. The ASAS-SN photometry shows no variation at the precision of the data. However, an analysis using a quasi-periodic Gaussian process (GP) suggested a 66\,$\pm$\,2\,d period signal in the data (see Fig.~\ref{fig:asas_lc}), albeit not with very high significance. We used the quasi-periodic GP kernel introduced by \cite{2017AJ....154..220F} of the form

\begin{equation*}
k_{i,j}(\tau) = \frac{B}{2+C}e^{-\tau/L}\left[\cos \left(\frac{2\pi \tau}{P_\textnormal{rot}}\right) + (1+C)\right] \quad,
\end{equation*}

\noindent where $\tau = |t_{i} - t_{j}|$ is the time-lag, $B$ and $C$ define the amplitude of the GP, $L$ is a timescale for the amplitude modulation of the GP, and $P_{\rm rot}$ is the period of the quasi-periodic modulations. For the fit, we considered that each instrument and pass band could have different values of $B$ and $C$, while $L$ and $P_{\rm rot}$ were left as common parameters. We considered wide uninformative priors for $B$, $C$ (log-uniform between $10^{-3}$ and $10^6$), $L$ (log-uniform between $10^{0}$\,d and $10^8$\,d), $P_{\rm rot}$ (uniform between 10\,d and 100\,d), and instrumental jitter (log-uniform between 10\,ppm and $10^6$\,ppm). The 3$\sigma$ upper limit implied by this latter fit is 300\,ppm, which demonstrates that \host{} is magnetically inactive with only very few starspots.  The photometric variability can be explained by a small starspot or group of starspots not exceeding approximately 2\,\% of the total area of the star assuming a star-spot temperature difference of 500\,K.  This is in agreement with the estimate of the RV amplitude that one could expect from rotational modulation following the prescription given by \citep{2012MNRAS.419.3147A}.  
Additionally, the H$\alpha$ activity indicator shows that \host{} is an inactive star \citep[][]{2018A&A...614A..76J}. Therefore, we conclude that the imprint of stellar activity signals in the collected CARMENES RVs is probably at the level of the measurement errors.

\section{Analysis and results}
\label{sec-analysis_and_results}

\subsection{Frequency analysis of radial velocities}
\label{subsec:gls}

\begin{figure}
    \centering
    \includegraphics[width=\linewidth]{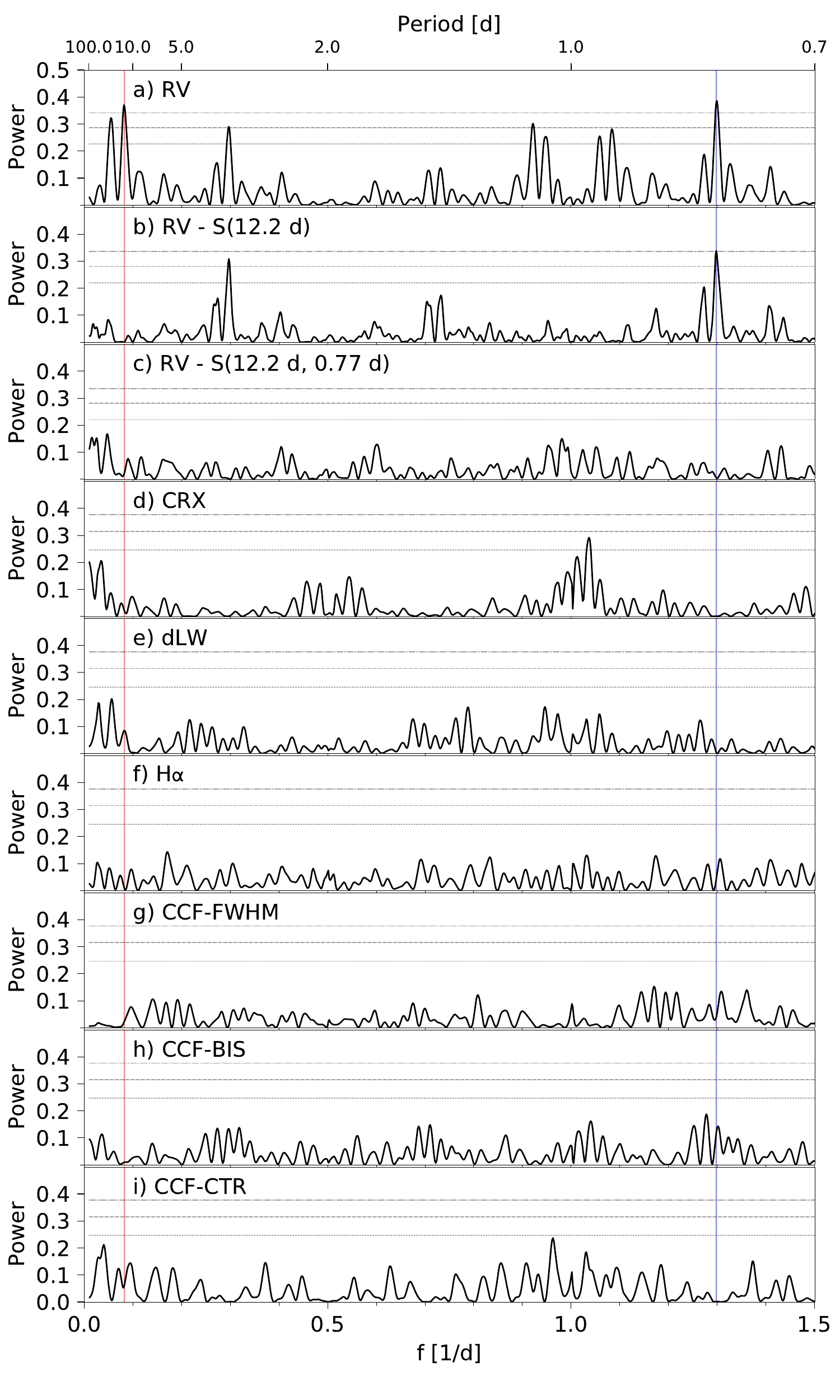}
    \caption{Generalised Lomb-Scargle periodograms for RVs of \host\ (a), their residuals (b) after fitting a sinusoid with period and phase corresponding to the transiting planet TOI-732.02 ($f_\mathrm{c} = 0.081\pm0.002\,\mathrm{d}^{-1}$, $P_\mathrm{c} = 12.3\pm0.3\,\mathrm{d}$), marked in red, and their residuals (c) after fitting two sinusoids with periods and phases corresponding to the transiting planets TOI-732.02 and TOI-732.01 ($f_\mathrm{b} = 1.298\pm0.002\,\mathrm{d}^{-1}$, $P_\mathrm{b} = 0.770\pm0.001\,\mathrm{d}$), marked in blue. Panels (d--i) show periodograms of the chromatic index, differential line width, H$\alpha$ index, cross-correlation function FWHM, bisector velocity span, and contrast, all of them derived only from CARMENES observations. Horizontal lines show the theoretical FAP levels of 10\,\% (short-dashed line), 1\,\% (long-dashed line), and 0.1\,\% (dot-dashed line) for each panel.
    \label{fig:gls_rv}}
\end{figure}

In order to search for the Doppler reflex motion induced by the transiting planets and unveil the presence of possible additional signals in our time-series RV data, we performed a frequency analysis of the RV measurements and their activity indicators. We calculated the generalised Lomb-Scargle (GLS) periodograms \citep{GLS} of the available time series and computed the theoretical 10\,\%, 1\,\%, and 0.1\,\% false alarm probability (FAP) levels (Fig.~\ref{fig:gls_rv}). The 70\,d time baseline of the RV measurements translates into a frequency resolution of 0.01428\,d$^{-1}$.

The GLS periodogram of the CARMENES data (Fig.~\ref{fig:gls_rv}a) shows two highly significant peaks ($\mathrm{FAP}<0.1\%$) at the orbital frequencies of the transiting planets, TOI--732.02 ($f_\mathrm{c} = 0.081\pm0.002\,\mathrm{d}^{-1}$, $P_\mathrm{c} = 12.3\pm0.3\,\mathrm{d}$) and TOI--732.01 ($f_\mathrm{b} = 1.298\pm0.002\,\mathrm{d}^{-1}$, $P_\mathrm{b} = 0.770\pm0.001\,\mathrm{d}$). The RV residuals after a joint fit with two circular orbits, fixed periods, and transit mid-times given by the \tess{} ephemerides showed no further significant peaks (Fig.~\ref{fig:gls_rv}c).

We also calculated periodograms for different activity indicators computed by SERVAL, namely the CRX (Fig.~\ref{fig:gls_rv}d), dLW (Fig.~\ref{fig:gls_rv}e), and H$\alpha$ index (Fig.~\ref{fig:gls_rv}f), and some indicators from the cross-correlation function such as FWHM, BIS, and CTR (Figs.~\ref{fig:gls_rv}g, \ref{fig:gls_rv}h and \ref{fig:gls_rv}i). No significant peaks were found except for some power with periods close to 1\,d in CRX, which are related to the sampling of the observations. There are no peaks in any activity indicator at the frequency of the transiting planets.

\subsection{Joint modelling of the light curves and RVs}
\label{subsec-analysis-joint_model}
We modelled the \tess{} light curve, the ground-based light curves, and the RVs simultaneously using \texttt{PyTransit} \citep{Parviainen2015}. The analysis followed the approach described in \citet{Parviainen2019,Parviainen2020}, namely we estimated any possible flux contamination from unresolved sources inside the photometry apertures in the \tess{} and ground-based photometry together with the planetary parameters.

The \tess{} dataset included in the analysis consisted of 2.4\,h windows of SAP light curves produced by the SPOC pipeline \citep{twicken:PA2010SPIE,Jenkins2016,2017ksci.rept....6M} centred around each individual transit centre normalised to the median per-window out-of-transit flux. The SAP light curves were chosen over the PDC light curves because the trends in the light curve are dominated by the photon noise (on 2.4\,h time scales), and because the PDC process removes the PDC-estimated flux contamination. The latter can introduce biases into our contamination estimation if the PDC contamination is overestimated, as we did not allow for negative contamination. By chance, the \tess{} dataset did not contain transits with overlapping windows. The ground-based photometry dataset included all the ground-based transit observations described in Section~\ref{ground_based_follow-up_observations}. We binned the MuSCAT2 photometry to a time cadence of one minute, but did not otherwise modify the data. The RV dataset was taken as it was, and included the 3.5\,m Calar Alto/CARMENES and Subaru/IRD usable RVs described in Section~\ref{sec-hrs}.

\begin{figure*}
    \centering
    \includegraphics[width=\linewidth]{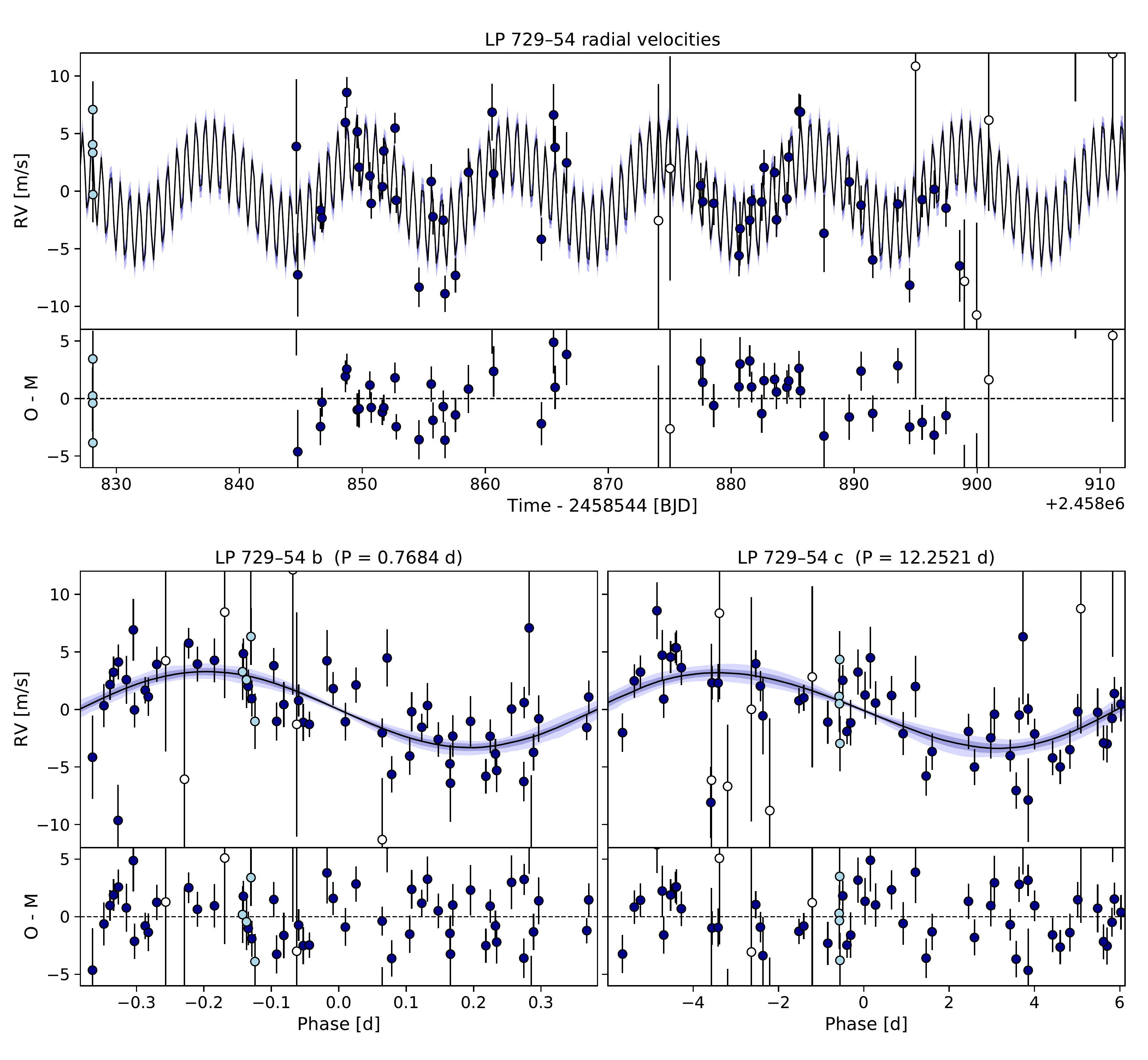}
    \caption{{\it Top panel:} CARMENES (black circles), IRD (light blue circles), and iSHELL (open circles) RV measurements along with the residuals of the median posterior joint fit model (black line) and the 68\,\%, 95\,\%, and 99\,\% central posterior limits (blue). {\it Bottom panels:} RVs phase-folded to the period of the two transiting planets ({\em left:} \planetb; {\em right:} \planetc).}
    \label{fig:joint_model_rvs}
\end{figure*}

\subsubsection{Parametrisation}
The joint model contained 207 free parameters, but most of them were linear coefficients used to model systematic trends in the ground-based light curves, and not of scientific interest. Of these 207 parameters, we used only 31 for describing the planets and their orbits, stellar limb darkening, and flux contamination.

\paragraph{Planet.}
Each planet and its orbit were parametrised by the orbital period, zero epoch, true planet-star area ratio, impact parameter, two parameters describing the eccentricity and argument of periastron, and RV semi-amplitude, as detailed in Table~\ref{tab:priors}. Here we made a distinction between the {apparent} and {true} planet-star area ratios. The {apparent area ratio} can be affected by flux contamination that leads to passband- and aperture-size-dependent variations in the apparent transit depth. However, the {true area ratio} stands for the uncontaminated geometric planet-star area ratio and, unlike the apparent area ratio, is independent of passband and photometry aperture size. In addition to these per-planet parameters, the stellar density was used to complete the description of the orbit.

\paragraph{Limb darkening.}
We parametrised the stellar limb darkening with the triangular parametrisation for the quadratic limb darkening model introduced by \citet{Kipping2013}, and constrained it using the code \texttt{LDTk} \citep{Parviainen2015b}. This yielded two parameters per passband, totalling 12 parameters for the six passbands for which we have photometric data.

\paragraph{Contamination.}
The \tess{} photometry was given an unconstrained contamination factor independent of the contamination for the ground-based observations, while the contamination in the ground-based observations was modelled using a physical model introduced by \citet{Parviainen2019}. This is because the \tess{} pixel size is significantly larger than the pixel size of the ground-based instruments used for the study, and thus we expected the contamination in the \tess{} data to be larger than in the ground-based photometry. In brief, the observed flux ($F_{\rm app}$), which can be used to determine the apparent planet-star radius ratio ($k_\mathrm{app}$) by fitting the transit model, in the light contamination model of \citet{Parviainen2019} is defined as a linear combination of the host and contaminant star fluxes (possibly from several contaminating sources). The contamination ($c$) is calculated for a set of passbands ($i$) given the passband transmission functions ($\mathcal{T}$), the effective temperatures of host ($T_\mathrm{eff,H}$), and contaminant stars ($T_\mathrm{eff,C}$), necessary to calculate relative fluxes of host ($F_\mathrm{H}$) and contaminant star ($F_\mathrm{C}$), and the level of contamination in some reference passband ($c_{0}$):
\begin{equation*}
c_i = \frac{F_\mathrm{C,i}}{F_\mathrm{H,i} + F_\mathrm{C,i}}.
\end{equation*}
By combining a contamination model with a transit model and taking into account that the transit depth scales linearly with the contamination factor, the true, uncontaminated radius ratio of a transiting planet ($k_\mathrm{true}$) can be calculated as:
\begin{equation*}
k_\mathrm{true} = k_\mathrm{app} / \sqrt{1-c}.
\end{equation*}

\paragraph{Trends and noise.}
We chose to use a simple linear model to explain the trends in the photometry. That is, the photometry (sans transit) was explained as a dot product of a baseline coefficient vector and a covariate vector. The ground-based light curves were observed with different instruments and the photometry was reduced with different photometry pipelines. Thus, the exact set of covariates included in the baseline model varied from light curve to light curve, but the airmass, $x$- and $y$-centroid shifts, and PSF FWHM were included whenever possible. The \tess{} photometry was given a constant baseline fixed to unity, because adding a per-transit model would increase the number of model parameters excessively.

We also chose to use a linear baseline model rather than to model the systematics as a GP because the latter approach would have added a significant layer of complexity to the analysis of such a heterogeneous dataset. Using a GP would require a separate kernel for each light curve source (since we have varying sets of covariates available), and the computation time would increase significantly.

\subsubsection{Joint modelling results.}
The analysis was carried out for two cases: ($a$) unconstrained contamination in the ground-based light curves, and ($b$) assuming no contamination in the ground-based light curves. The contamination in the \tess{} photometry was unconstrained in both cases, and independent of the ground-based contamination.

The first case, ($a$) was used to determine whether or not the ground-based photometry contained additional flux from any unresolved source. The analysis excluded significant contamination from any source of different spectral type than the host star in the ground-based observations (i.e. only passband-independent contamination was allowed, which would require the contaminating source to have the same spectral type as the host star). 

Since all the additional observational data rule out an almost identical nearby star, we assumed that the ground-based photometry did not contain significant contamination, in agreement with the results of the ground-based follow-up presented in Sect.~\ref{ground_based_follow-up_observations}. Therefore, we chose the parameter posteriors from the second case as our final parameter estimates, and present them here. The parameter estimates for the first case are available from {\tt GitHub} with the rest of the analyses.

\begin{figure}
    \centering
    \includegraphics[width=\columnwidth]{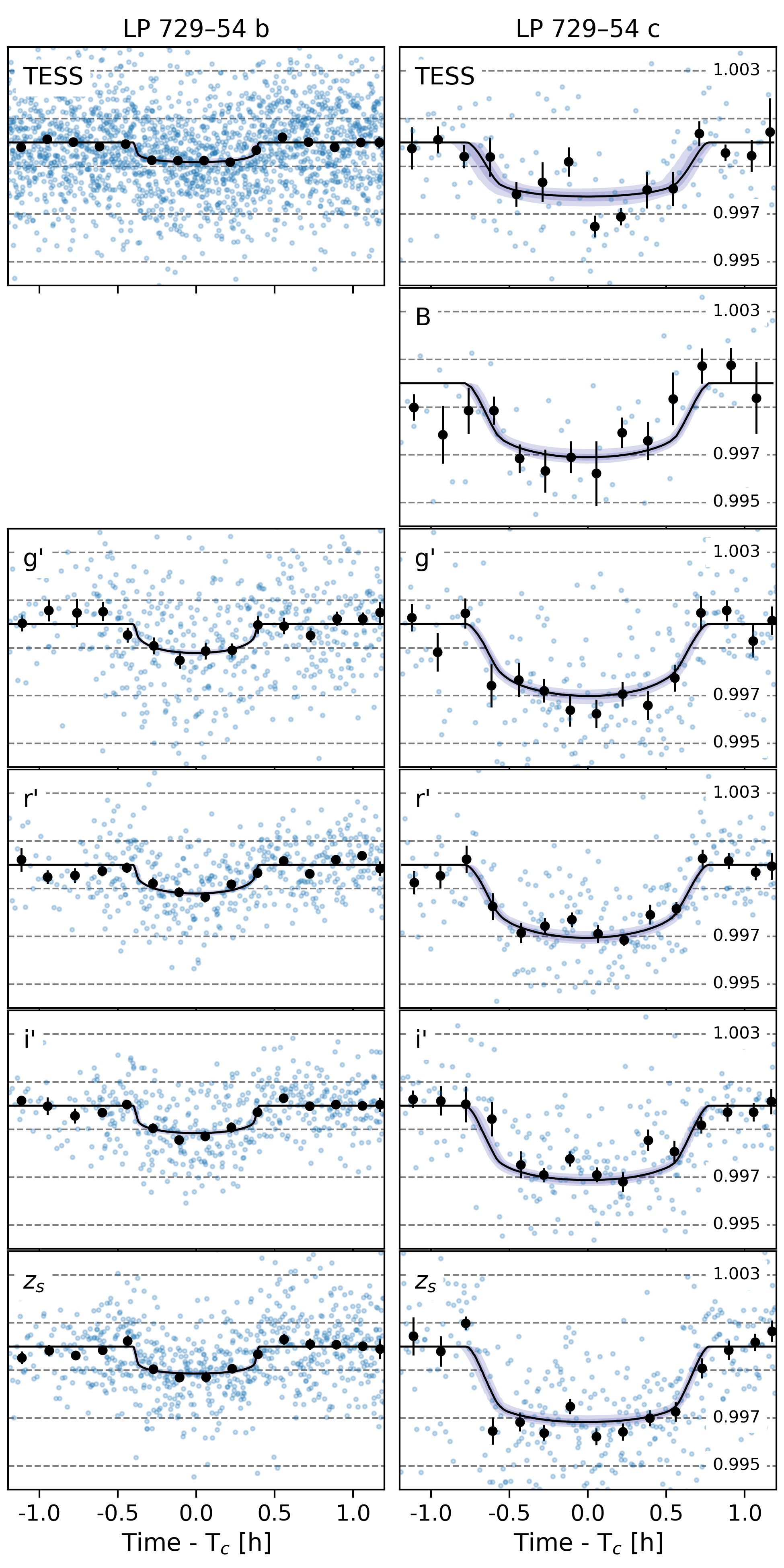}
    \caption{Combined and phase-folded transits of \planetb{} and c for each passband. The blue points show the original photometry with the median baseline model removed, the black dots with error bars show the photometry binned to 10\,min resolution, the black line shows the median posterior model, and the dark and light shaded areas show the 68\,\% and 95\,\% model posterior percentile limits, respectively.  }
    \label{fig:joint_model_lcs}
\end{figure}

Finally, we report the model posterior parameter estimates in Table~\ref{table:posterior_parameters}. The posterior model is shown in Figs.~\ref{fig:joint_model_rvs}~and~\ref{fig:joint_model_lcs}, and parameter posteriors are shown for selected parameters in Figs.~\ref{fig:corner_01} and \ref{fig:corner_02} in Appendix~\ref{sec:joint_modelling_posteriors}.

\begin{table*}
        \centering
        \normalsize
        \caption{Posterior estimates for the stellar and planetary parameters from the combined analysis.}
        \begin{tabular}{@{}lllcc@{}}
                \toprule\toprule
                Quantity & Notation & Unit & \planetb{} & \planetc{} \\
                \midrule
                \noalign{\smallskip}
                \multicolumn{5}{l}{\emph{~~~~~Orbit}} \\
                \noalign{\smallskip}
                Transit epoch & $T_0$ & BJD & \pvtcb & \pvtcc \\
                Orbital period & $P$ & d & \pvpb & \pvpc \\
                Eccentricity & $e$ & ... & \pveb & \pvec \\
                Argument of periastron & $\omega$ & deg & \pvwdegb & \pvwdegc \\
                Transit duration & $T_{14}$ & h &\pvtdurb & \pvtdurc \\
                \noalign{\smallskip}
                \multicolumn{5}{l}{\emph{~~~~~Relative properties}} \\
                \noalign{\smallskip}
                Radius ratio &$k = R_p / R_\star$ & ... & \pvkb & \pvkc \\
                Scaled semi-major axsi &$a_\mathrm{s} = a_p / R_\star$ & ... & \pvasb & \pvasc \\
                Impact parameter & \ppb & ... & \pvbb & \pvbc \\
                \noalign{\smallskip}
                \multicolumn{5}{l}{\emph{~~~~~Absolute properties}} \\
                \noalign{\smallskip}
                Planet radius & \ppr & \Rea  &  \pvrpb & \pvrpc \\
                Planet mass & \ppm  & \Mea & \pvmpb & \pvmpc \\
                Mean density & \ppden & \ppdenu & \pvrhopb & \pvrhopc\\
                Semi-major axis& \ppa & \ppau &  \pvab & \pvac \\
                Gravitational acceleration & \ppgra & \ppgrau & \pvgpb & \pvgpc \\
                Equilibrium temperature & \ppteq & \pptequ & \pvteqpb & \pvteqpc \\
                Inclination & \ppi & deg & \pvincdegb & \pvincdegc\\
                Insolation & \ppins & \ppinsu & \pvinspb & \pvinspc \\
                \bottomrule       
        \end{tabular}
        \label{table:posterior_parameters}  
\end{table*}

\section{Discussion}
\label{sec-discussion}
\subsection{\texorpdfstring{\planetb}{LTT\,3470\,b} and\,c: two planets in the same system straddling the radius gap}

\begin{figure}
    \centering
    \includegraphics[width=\linewidth]{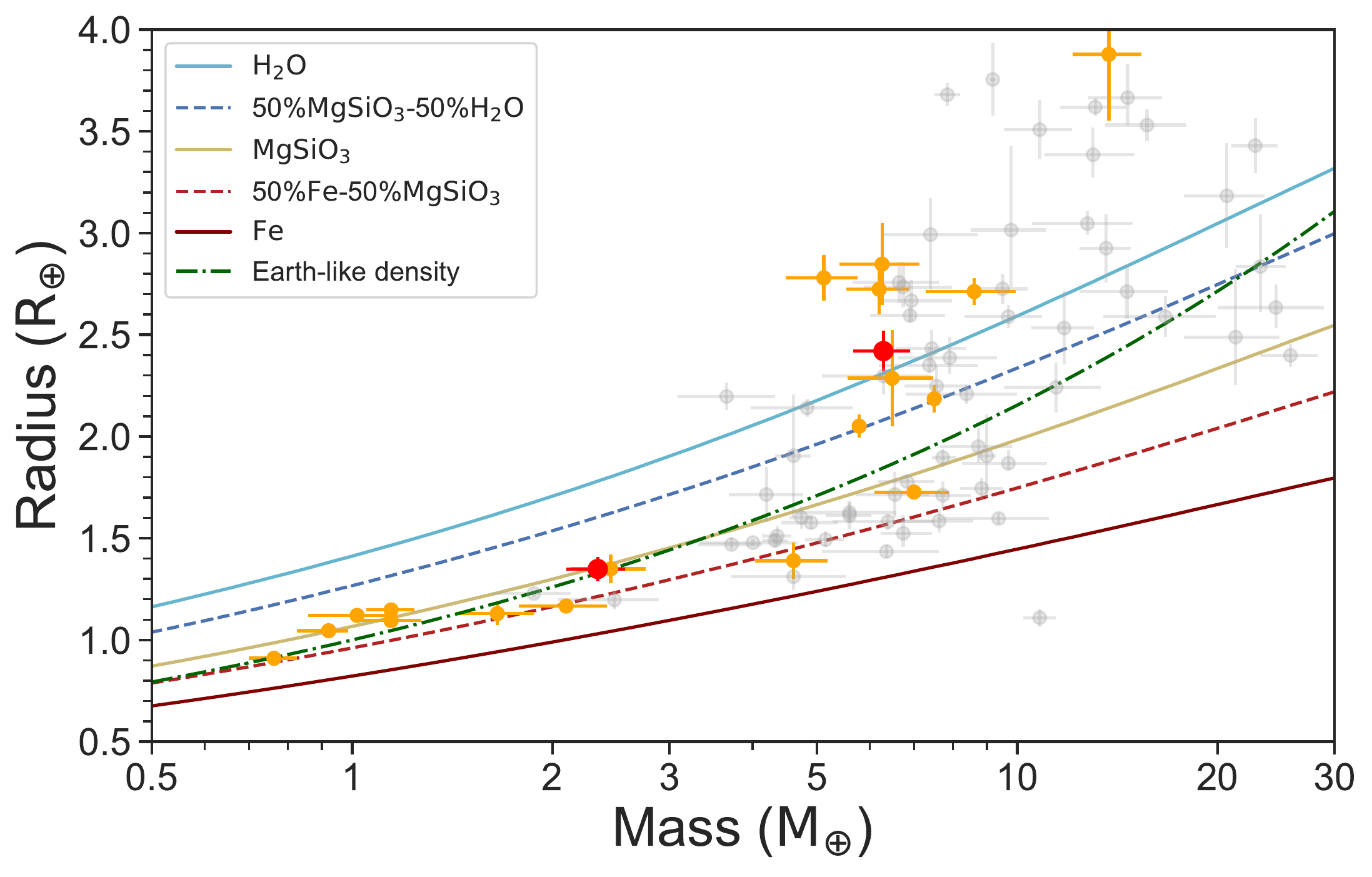}
    \caption{Mass-radius diagram for all planets with mass and radius measurement better than 20\,\% (from the TEPCat\protect\footnotemark\ database of well-characterised planets; \citealt{2011MNRAS.417.2166S}). M-dwarf host planets are shown in orange, {\planetb} and {\planetc} with red dots. Theoretical models \citep{2016ApJ...819..127Z} are overplotted using different lines and colours.
    \label{plot_mr}}
\end{figure}

The \host{} system consists of two small transiting planets. The ultra-short-period planet \planetb{} (\pbporb\,$\approx$\,\pbporbva) has a radius of \pbr\,=\,\pbrv{} and a mass of \pbm\,=\,\pbmv{}, yielding a mean density of \pbden\,=\,\pbdenv. The outer planet \planetc{} (\pcporb\,$\approx$\,\pcporbva) has a radius of \pcr\,=\,\pcrv{} and a mass of \pcm\,=\,\pcmv{}, yielding a mean density of \pcden\,=\,\pcdenv. This is the first planetary system around an M dwarf with two planets located on opposite sides of the radius gap \citep{2017AJ....154..109F,2018AJ....156..264F} that separates super-Earths from sub-Neptunes.

\addtocounter{footnote}{-1}
The positions of \planetb{} and \planetc{} on the mass-radius diagram are shown in Fig. 9 in comparison to the sample of small transiting planets (\ppr\,$\leq$\,4\,\Rea) whose masses and radii have been derived with a precision better than 20\,\%\footnote{\url{http://www.astro.keele.ac.uk/jkt/tepcat/}} . The bulk densities of the two planets are significantly different, and their positions in the mass--radius diagram indicate substantially different compositions. Planet \planetb{} is compatible with an Earth-like bulk composition, ranging from 50\,\% silicate and 50\,\% iron to 100\,\% silicate. On the other hand, \planetc{} has a bulk density consistent with a volatile-dominated world. 

\begin{figure}
\centering
\includegraphics[width=\linewidth]{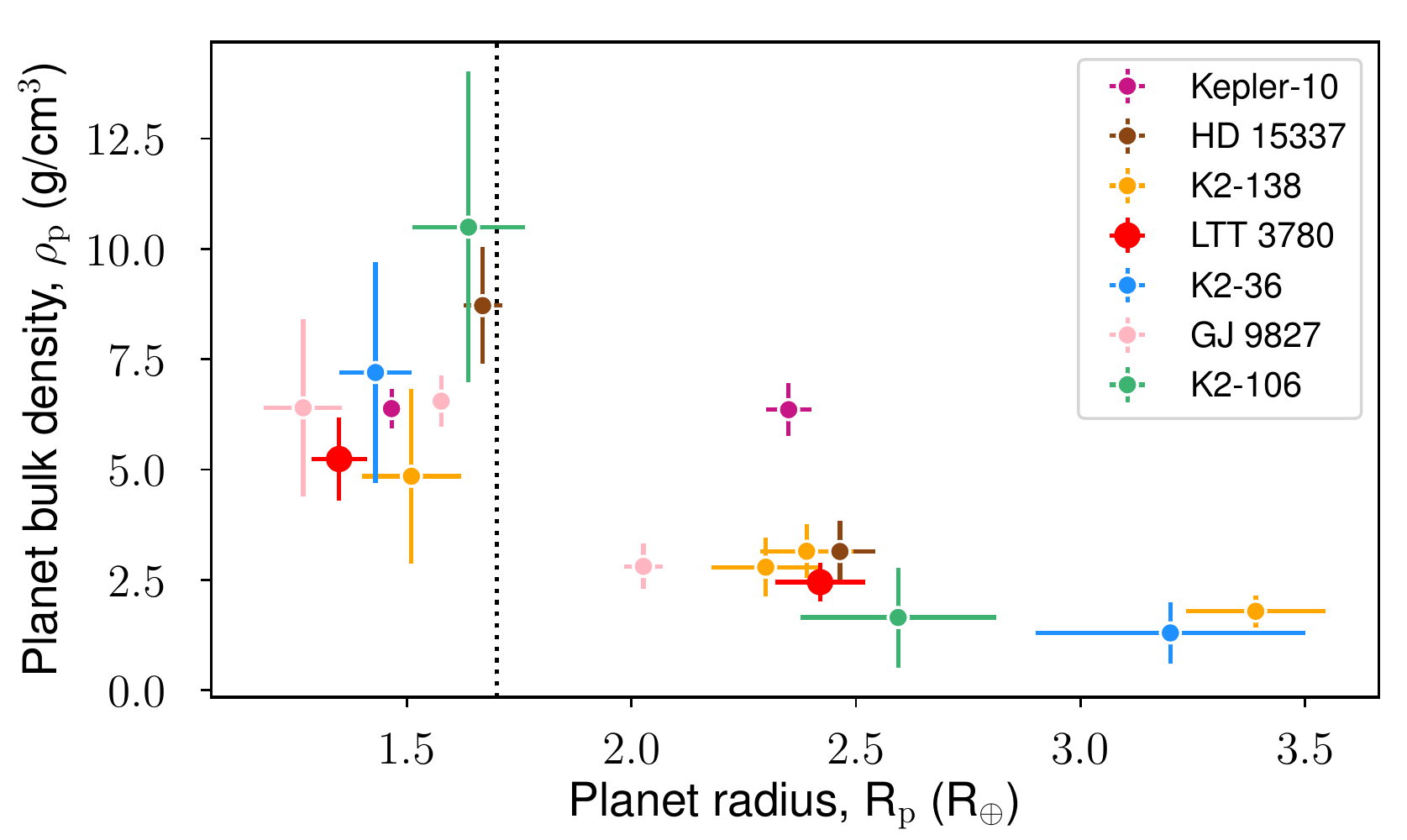}
\caption{Radius--density diagram for multi-planet systems with planets on both sides of the radius gap. Different colours represent the different planetary systems, with the \host{} planets marked and labelled in red. The vertical dotted line marks the centre of the radius gap at 1.7\,\Rea.
\label{psrg-pr-pd}}
\end{figure}

The architecture of the \host{} planetary system is consistent with those of other systems hosting two or more small, close-in  planets located on opposite sides of the radius gap: Kepler-10~bc \citep{2014ApJ...789..154D}, K2-106~bc \citep{2017AJ....153..271S,2017A&A...608A..93G}, HD~3167~bc \citep[K2-96~bc;][]{2017AJ....154..122C,2017AJ....154..123G}, GJ~9827~bcd \citep[K2-135~bcd;][]{2017AJ....154..266N,2018A&A...618A.116P}, K2-138~bcdef \citep{2018AJ....155...57C,2019A&A...631A..90L}, HD~15337~bc \citep[TOI-402~bc;][]{2019ApJ...876L..24G,2019A&A...627A..43D}, and K2-36~bc \citep{2019A&A...624A..38D}. In all these systems, the close-in planets have smaller radii and higher mean densities, consistent with a rocky terrestrial composition, and the outer planets have larger radii and lower mean densities, suggesting that they are composed of rocky cores surrounded by light, hydrogen-dominated or water envelopes (see Fig.~\ref{psrg-pr-pd}). This result agrees with current theoretical scenarios that explain the existence of the radius gap by atmospheric escape \citep{2017ApJ...847...29O,2018ApJ...853..163J}.

All previously-known stars with small planets located on opposite sides of the radius gap have warmer effective temperatures and higher masses than \host. Except for GJ~9827\footnote{GJ~9827 should have been named BD--02~5958, as the third and last Gliese/Gliese-Jahreiss catalogue counted only from GJ~1 to GJ~4388 \citep{1991adc..rept.....G}.}, host stars effective temperatures and masses range between 4920\,K and 5810\,K and 0.80\,M$_\odot$ and 0.93\,M$_\odot$, respectively. The previous smallest and coolest such star host, GJ~9827, has $T_{\rm eff} \approx$ 4260\,K and $M \approx$ 0.66\,M$_\odot$, consistent with its late K spectral type \citep{1974ApJS...28....1J,1985ApJS...59..197B,1986AJ.....92..139S}.
However, \host{} is significantly cooler ($T_{\rm eff} \approx$ 3360\,K) and, correspondingly, less massive ($M \approx$ 0.38\,M$_\odot$).

\subsection{The ultra-short period planet \texorpdfstring{\planetb}{LTT\,3470\,b}}
With orbital period of \pbporb\,$\approx$\,\pbporbva{}, \planetb{} belongs to the population of ultra-short period planets (USPs) and joins the relatively small group of transiting USPs with masses precisely measured via radial velocities and, hence, determined bulk densities. Taking into account the density of \planetb\ (\pbden\,=\,\pbdenv), this USP has probably undergone significant evolution and lost its primary, hydrogen-dominated atmosphere.
The low luminosity of \host, as expected from its late spectral type, is counterbalanced by the short semi-mayor axis of \planetb{}, which leads to a strong planetary insolation (\pbins = \pvinspb\,\ppinsu). 

The stellar properties of \host{} and GJ~1151 (M4.5\,V) are very similar \citep{2018A&A...615A...6P}, both being relatively quiet stars. Although \host{} is  more distant from Earth and about half a magnitude fainter, it is worth searching for low-frequency radio emission from \host{} possibly generated by the interaction of the star's magnetospheric plasma with its USP, as suggested in the case of GJ~1151 \citep{2020NatAs.tmp...30V}.

\subsection{System architecture}
The architecture of the \host{} system is analogous to that of Kepler-10, which has a (slightly larger and commensurately more massive) rocky planet with an orbital period about 10\,\% longer than that of \planetb{} \citep{2011ApJ...729...27B}, and an outer planet with orbital period of 45\,d, whose radius is within a few percent of that of \planetc{} \citep[the mass of Kepler-10 c is poorly-constrained;][]{2011ApJS..197....5F,2016ApJ...819...83W}, and whose orbit is also inclined by $\sim$5--6\,deg with respect to its inner ultra-short-period sibling. 
Multi-transiting planetary systems have few planets with orbital periods less than 1.6\,d \citep{2014ApJ...784...44L}, and the high inclination of these two well-studied multi-planet systems with USPs (the inclinations of most other USP in {\em Kepler} multi-planet systems are not well constrained) supports the hypothesis that this paucity results at least in part from typical USPs being more highly inclined with respect to their planetary companions than the 1--2\,deg value typical for {\em Kepler} multi-planet systems found by \cite{2014ApJ...790..146F}.

\subsection{Atmospheric scenarios for \texorpdfstring{\planetc}{LTT\,3470\,c}}

\begin{figure*}
    \includegraphics[width=\textwidth]{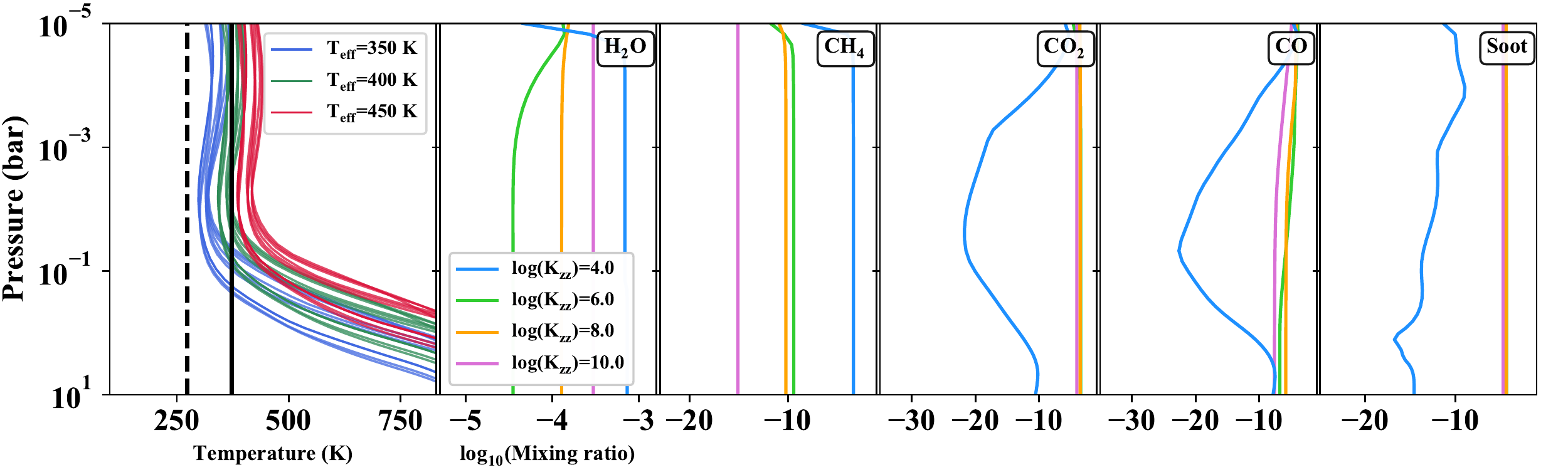}
    \caption{{\it Left panel:} Simulated temperature structures of {\planetc} assuming a substantial primary atmosphere. Blue, green, and red curves represent the temperature profiles at different effective temperatures of 350, 400, and 450\,K, respectively. Different profiles within each group are caused by different atmospheric metallicity and C/O ratio. There are 27 temperature structures in total. The dashed and solid black vertical lines mark 273.15\,K (0\degree C) and 373.15\,K (100\degree C) for reference. 
    {\it Remaining panels:} Examples of atmospheric abundances of \ce{H2O}, \ce{CH4}, \ce{CO2}, \ce{CO}, and soot (haze particles), for solar metallicity and C/O ratio, and effective temperature of 400\,K, resulting from the photo-chemical simulations at different vertical mixing strengths.}
\label{fig:TPs}
\end{figure*}

\begin{figure*}
    \includegraphics[width=\textwidth]{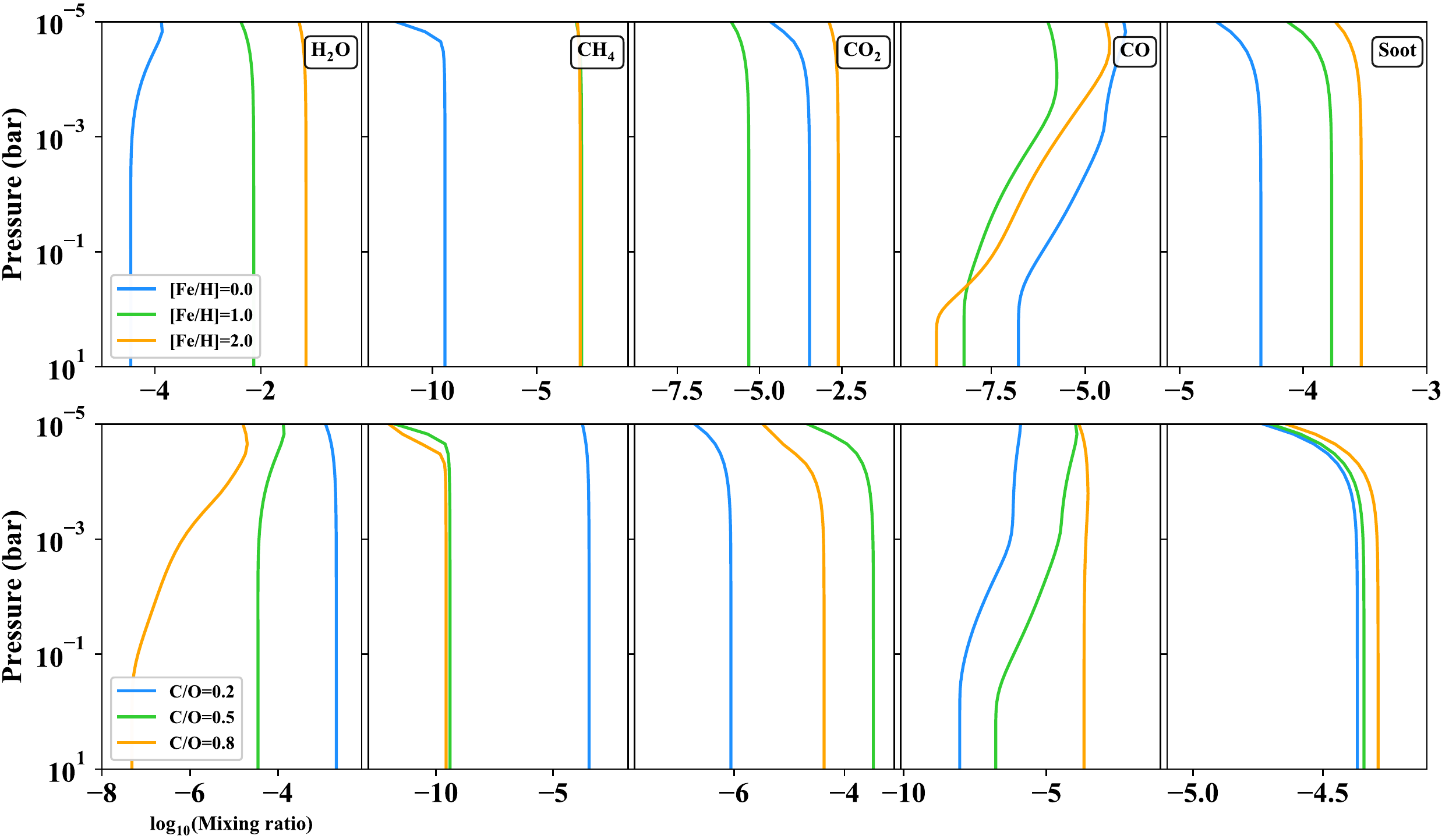}
    \caption{Abundances of several atmospheric constituents at $T_{\rm eff}=400$\,K.
    {\it Top panels:} Solar C/O ratio, and K\textsubscript{zz}=10$^6$\,cm$^2$\,s$^{-1}$.
    {\it Bottom panels:} Solar metallicity and C/O ratios varying from 0.2 to 0.8.}
\label{fig:abunds}
\end{figure*}

The estimated radii of \planetb{} and c provide a unique opportunity to study the mechanisms that shape the Fulton gap \citep{2017AJ....154..109F}. While the proximity of \planetb{} to its host star and its relatively small radius suggest the challenging nature of maintaining an atmosphere on this planet, characterisation of the atmosphere of \planetc{} will shed light on the nature of the dominant atmospheric processes in action on this planet. While \planetc{} is not as appealing a target as TRAPPIST planets \citep[TSM $\in$ 20--45;][]{2017Natur.542..456G,2018A&A...613A..68G} or LHS\,1140\,b \citep[TSM$\sim$65;][]{2017Natur.544..333D}, its transmission spectroscopy metric ($\mathrm{TSM}$) defined by \citet{2018PASP..130k4401K} for {\em JWST}/NIRISS is $\sim$ 122, which is above the cutoff of TSM = 90 suggested for atmospheric characterisation of planets with radii 1.5\,<\,\ppr\,<\,2.75\,\Rea{} with JWST/NIRISS.

In order to explore possible atmospheric scenarios of \planetc, we followed the approach outlined in \citet{2019ApJ...883..194M}, assuming that the planet sustains a substantial primary atmosphere. First we modelled the atmosphere self-consistently over a wide range of parameters using {\tt petitCODE} \citep{2015ApJ...813...47M,2017A&A...600A..10M}. Given the uncertainties on the estimated equilibrium temperature of \planetc{} and possible missing feedback mechanisms in our self-consistent simulations, we chose the effective temperature as a free parameter with values of 350, 400, and 450\,K (the equilibrium temperature derived for \planetc{} is \pvteqpc\,K). 
Since the interior heat budget of exoplanets is not well understood, we assumed an interior heat budget contribution similar to that of Earth, i.e. $\sim$0.027 \citep[e.g.][]{archer_global_2011}. Slight deviations of the interior heat budget from this value do not change the results significantly. The atmospheric metallicity of planets seems to increase for less massive planets, both in the Solar System \citep[e.g.][]{2019ApJ...873...32M} and beyond \citep{wakeford_hat-p-26b:_2017}. Thus, we assumed three different metallicities, namely 1$\times$, 10$\times$, and 100$\times$ solar metallicity, similar to those used by \citet{luque_planetary_2019}. In addition, we also investigated the role of the carbon-to-oxygen ratio (C/O) in the atmosphere of \planetc, assuming C/O = 0.2 (sub-solar), 0.5 ($\sim$solar), and 0.8 (super-solar). Altogether, 27 self-consistent temperature structures were calculated (Fig.~\ref{fig:TPs}). No Earth-like cloud formation was included in our models.

In the next step we used these temperature structures as the input of our photochemical model \citep[\texttt{ChemKM};][]{2019ApJ...883..194M,2020arXiv200103668M} to estimate abundances of atmospheric constituents. Studying the atmosphere of this planet required a validated chemical network over the assumed equilibrium temperatures. We therefore used the \citet{hebrard_neutral_2012} full kinetic network (including 788 reactions and 135 H-C-O-N bearing species) and an updated version of their ultraviolet absorption cross-sections and branching yields. While the formation of hydrocarbon-based hazes is not fully understood \citep[e.g.][]{horst_haze_2018}, it is believed that these processes start with the photolysis of haze precursor molecules such as \ce{CH4}, \ce{C2H2}, HCN, and \ce{C6H6} \citep[e.g.][]{2020arXiv200103668M}. The chosen chemical network included all these precursor molecules and represented the haze particles collectively as one constituent called `soot' \citep[e.g.][]{lavvas_aerosol_2017}. 
This soot included \ce{C8H6}, \ce{C8H7}, \ce{C10H3}, \ce{C12H3}, \ce{C12H10}, \ce{C14H3}, \ce{C2H4N}, \ce{C2H3N2}, \ce{C3H6N}, \ce{C4H3N2}, \ce{C4H8N}, \ce{C5HN}, \ce{C5H3N}, \ce{C5H4N}, \ce{C5H6N}, \ce{C9H6N}, \ce{C3H3O}, \ce{C3H5O}, \ce{C3H7O}, and \ce{C4H6O}, following the convention outlined by \citet{hebrard_neutral_2012}.

\begin{figure*}
\includegraphics[width=\textwidth]{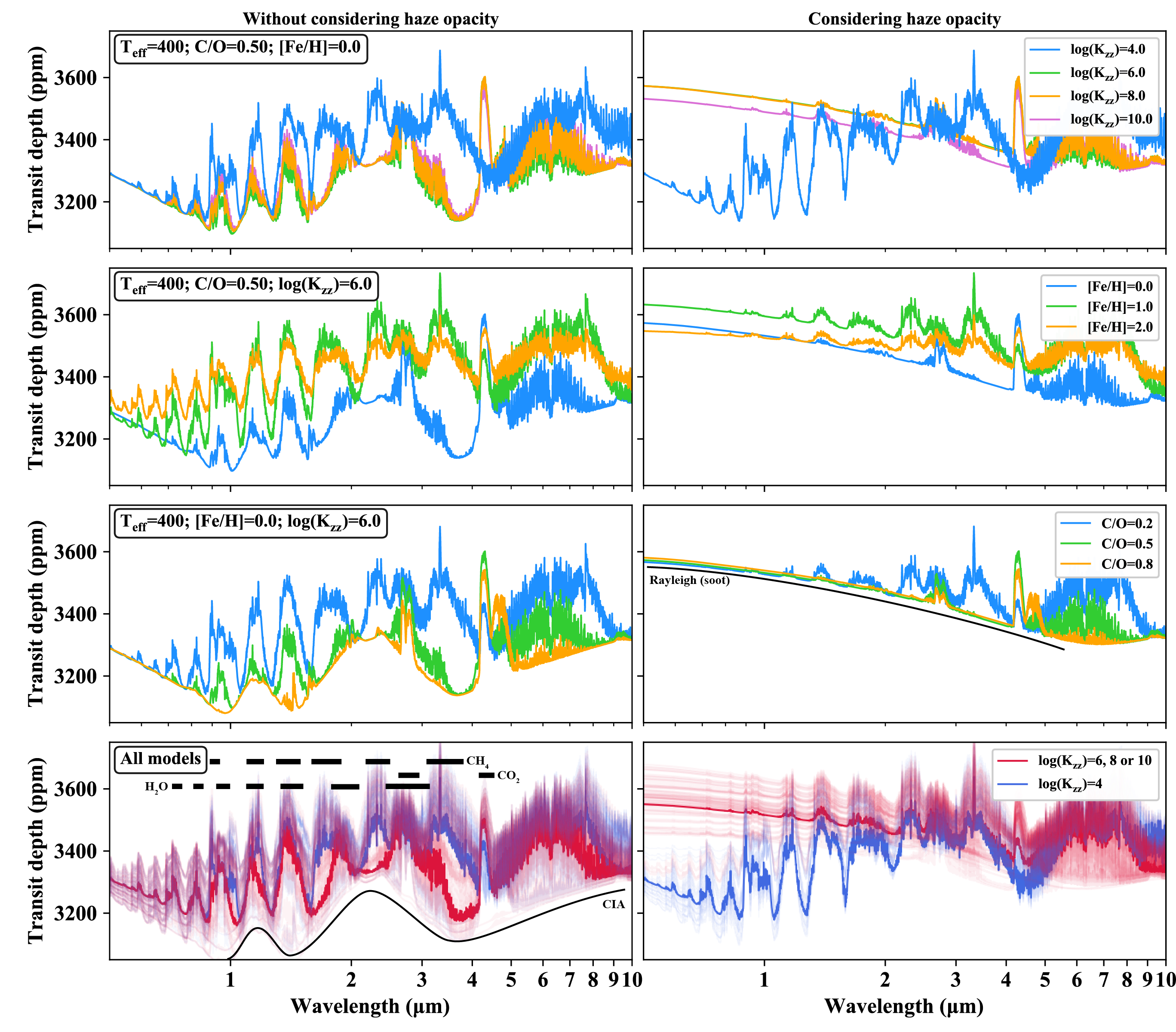}
\caption{Synthetic transmission spectra of \planetc. {\em Left panels:} Transmission spectra calculated without taking into account the opacity contribution of haze particles (soot). The top-left panel shows the variation of spectra with the strength of vertical mixing, corresponding to the cases in Fig.~\ref{fig:TPs}. The two middle panels also correspond to the two cases in Fig.~\ref{fig:abunds}. The bottom-left panel represents the transmission spectra of all 108 models in the two categories of strong (10$^6$, 10$^8$, or 10$^{10}$\,cm$^2$s$^{-1}$) and weak (10$^4$\,cm$^2$s$^{-1}$) vertical mixing. {\em Right panels:} Same as left panels, but with opacity of haze particles taken into account.}
\label{fig:spectra_tr}
\end{figure*}

\addtocounter{footnote}{-1}
The temperature of GJ~667C, another M-dwarf exoplanet host, is estimated to be around 3330\,K \citep{neves_Metallicity_2014}, similar to that of \host. Consequently, we estimated the flux of \host{} in the range of X-ray to optical wavelengths using GJ~667C data, obtained from the MUSCLES database \citep{france_muscles_2016}. In addition to photolysis, we considered the effect of vertical mixing in the photochemical simulations by considering four values, 10$^4$, 10$^6$, 10$^8$, and 10$^{10}$\,\cmpss, covering a wide range of possibilities from terrestrial values to tidally locked gaseous planets. In total, 108 photochemical models were calculated\footnote{The atmospheric temperature structures, abundances and transmission and emission spectra of these 108 models are publicly available at \url{www.mpia.de/homes/karan/}.}.

Examples of \ce{H2O}, \ce{CH4}, \ce{CO2}, \ce{CO}, and soot (haze particles) abundances are shown in Fig.~\ref{fig:TPs}, assuming an effective temperature of 400\,K, solar metallicity and solar C/O. In general, chemical depletion of \ce{H2O} and \ce{CH4} are noticeable at strong vertical mixing conditions, i.e. 10$^6$, 10$^8$ and 10$^{10}$\,\cmpss, while \ce{CO2}, \ce{CO}, and soot particles show an enhancement in abundance under such conditions. Figure~\ref{fig:abunds} isolates the effect of metallicity (upper panels) and C/O ratio (lower panels) on the abundance of \ce{H2O}, \ce{CH4}, \ce{CO2}, \ce{CO}, and soot. For these illustrated examples, we assumed an effective temperature of 400\,K and a vertical mixing strength (K\textsubscript{zz}) of 10$^6$\,cm$^2$\,s$^{-1}$. In general, increasing the metallicity would enhance the production of most species at most altitudes under these conditions. However, given the highly non-linear nature of atmospheric feedback, some species could behave differently; see e.g.\ CO in the upper panel of Fig.~\ref{fig:abunds}. 
The C/O ratio is modified by changing the oxygen elemental abundance and keeping the carbon elemental abundance the same. This represents a scenario in which gas or planetesimals accrete onto a forming planet with different water content, while the carbon content assumed to remain the same.
Hence, variations of the C/O should not change methane abundances very much, as it has no oxygen compound. However, as a function of  the dependency of the chemical formation pathway of a constituent on the formation of oxygen-bearing species, as well as their radiative feedback, the abundance of hydrocarbons could change with the C/O ratio. The lower panels of Fig.~\ref{fig:abunds} illustrate such an example, where \ce{CH4} remains mostly insensitive to the C/O ratio at C/O $\geq$ 0.5, but a C/O of 0.2 results in an enhanced abundance of methane. This unexpected methane production (or lack of methane depletion) is largely caused by how disequilibrium processes, namely photo-dissociation and atmospheric mixing, act on this planet. In addition, a variation of the C/O ratio affects the \ce{H2O} abundance significantly, as is shown in the lower-left panel of Fig.~\ref{fig:abunds}.

We find that, due to the relatively low temperature of \planetc, any vertical mixing stronger than 10$^4$\,cm$^2$\,s$^{-1}$ could strongly quench the abundance of most species. This appears as a nearly constant abundance profile for any given species at such K\textsubscript{zz}. Figure~\ref{fig:TPs} illustrates examples for this at three different values of K\textsubscript{zz}. This is particularly important for haze particles because under these circumstances their abundance becomes very significant at all altitudes and  can therefore affect the spectra by obscuring the atomic and molecular features in the optical and NIR ranges.

Figure 13 shows synthetic transmission spectra of the above-mentioned cases calculated using {\tt petitRADTRANS}
\citep{2019A&A...627A..67M}.  
We considered two cases, one without haze opacity, and the other with haze opacity assuming a mono-disperse particle size distribution with an effective particle radius of 50\,nm \citep[e.g.][]{Tomasko2005,Trainer18035}. The change in the C/O ratio had only a small effect compared to other parameters, where it mainly affected the spectral significance of \ce{H2O}. On the other hand, a combination of low metallicity and strong vertical mixing could change the transmission spectra significantly. Under such conditions, the atmosphere becomes methane-depleted. The case of methane-depletion at K\textsubscript{zz} $>$ 10$^4$~cm$^2$s$^{-1}$ and [Fe/H] = 0.0 (solar metallicity) is shown in the top panels of Fig.~\ref{fig:spectra_tr}. Similarly, at K\textsubscript{zz} = 10$^6$\,cm$^2$\,s$^{-1}$, solar metallicity results in the depletion of methane, second row from top in Fig.~\ref{fig:spectra_tr}. Another key feature is the presence of \ce{CO2} at around 4.3\,\microm{} with a spectral difference above 200\,ppm for most cases with strong mixing. Prominent spectral features (associated with \ce{CH4}, \ce{H2O}, and \ce{CO2}) are marked in the lower left panel of Fig.~\ref{fig:spectra_tr}. The collision-induced absorption (CIA) continuum is also shown in the same panel.

The right panels of Fig.~\ref{fig:spectra_tr} show the same cases as the left panels, except that they include the opacity of haze particles in the synthetic transmission spectra of \planetc. The significance of this opacity in the spectra is evident. As mentioned, haze production is mainly associated with strong vertical mixing; see top right and bottom right panels. Including the haze opacity makes the detection of methane-depleted atmospheres more challenging. However, the \ce{CO2} feature at 4.3\,\microm{} remains a distinct and key feature for the detection of atmospheric features in \planetc{} in the NIR. Our simulations suggest that \jwst{} instruments would be able to detect the \ce{CO2} spectral feature, as their noise floor is much lower than the spectral significance of this feature \citep{greene_characterizing_2016}.
 
In summary, despite many degeneracies between the synthetic spectra, looking for the presence of \ce{H2O}, \ce{CH4}, \ce{CO2}, and haze particles (Rayleigh scattering) with future facilities such as \jwst could hint at the type and significance of the dominant atmospheric processes on \planetc.

On a final note, the extension of simulations to 3D models is likely to favour more pronounced cloud formation and stronger disequilibrium processes on the night side and their extension to at least the morning limb \citep[e.g.][]{2020arXiv200103668M}. Hence, 3D simulations are needed to evaluate the significance of these processes. Moreover, further analyses are needed to study the processes by which the primary atmosphere of both \planetb{} and c could be removed, and a secondary atmosphere could be formed and maintained.

\subsection{An independent analysis of the \host{} system by HARPS and HARPS-N}
Following the announcement of the planet candidates TOI--732.01 and 02 in 2019 May, multiple precision RV instrument teams began working toward the mass characterisation of these potential planets. The present study presents the subset of those efforts from CARMENES, but we are aware that the joint HARPS and HARPS-N team has also submitted a paper presenting their own RV time series and analysis \citep{2020AJ....160....3C}. Although the submissions of these complementary studies were coordinated between the two groups, their respective data, analyses, and write-ups were intentionally conducted independently.

\section{Conclusions}
\label{sec-conclusions}

We report here the discovery and detailed characterisation of a planetary system around the bright M dwarf \host{}, composed of two small transiting planets straddling the radius gap. Transit signals of both planets are detected in the \tess{} photometry and confirmed by ground-based facilities. These ground-based photometric observations of \host{} allow us to confirm the planetary nature of the candidates and determine the true radii of planets thanks to aperture photometry uncontaminated by close-in stars. Prompt RV follow-up with the CARMENES spectrograph provides a precise mass determination of both planets.

\planetb{} is a very hot (\ppteq\,=\,\pbteqv), ultra-short period (\pbporb\,$\approx$\,\pbporbva) planet that is slightly larger in size than the Earth (\pbr\,=\,\pbrv) but has an equal  bulk density. With radius \pcr\,=\,{\pcrv}, \planetc{} is located on the opposite side of the radius gap, exactly at the peak of the sub-Neptune distribution that is believed to be a population of gas-dominated planets. With an equilibrium temperature of \pcteq\,=\,\pcteqv, {\planetc} is an excellent target for atmospheric characterisation with the upcoming \jwst.

\begin{acknowledgements}
CARMENES is an instrument for the Centro Astron\'{o}mico Hispano-Alem\'{a}n de Calar Alto (CAHA, Almer\'{\i}a, Spain). CARMENES is funded by the German Max-Planck-Gesellschaft (MPG), the Spanish Consejo Superior de Investigaciones Cient\'{\i}ficas (CSIC), the European Union through FEDER/ERF FICTS-2011-02 funds, and the members of the CARMENES Consortium (Max-Planck-Institut f\"{u}r Astronomie, Instituto de Astrof\'{\i}sica de Andaluc\'{\i}a, Landessternwarte K\"{o}nigstuhl, Institut de Ci\`{e}ncies de l'Espai, Institut f\"{u}r Astrophysik G\"{o}ttingen, Universidad Complutense de Madrid, Th\"{u}ringer Landessternwarte Tautenburg, Instituto de Astrof\'{\i}sica de Canarias, Hamburger Sternwarte, Centro de Astrobiolog\'{\i}a and Centro Astron\'{o}mico Hispano-Alem\'{a}n), with additional contributions by the Spanish Ministry of Economy, the German Science Foundation through the Major Research Instrumentation Programme and DFG Research Unit FOR2544 "Blue Planets around Red Stars", the Klaus Tschira Stiftung, the states of Baden-W\"{u}rttemberg and Niedersachsen, and by the Junta de Andaluc\'{\i}a.

This paper includes data collected by the \tess{} mission. Funding for the \tess{} mission is provided by the NASA Explorer Program. We acknowledge the use of public TOI Release data from pipelines at the \tess{} Science Office and at the \tess{} Science Processing Operations Center. Resources supporting this work were provided by the NASA High-End Computing (HEC) Program through the NASA Advanced Supercomputing (NAS) Division at Ames Research Center for the production of the SPOC data products. This research has made use of the Exoplanet Follow-up Observation Program website, which is operated by the California Institute of Technology, under contract with the National Aeronautics and Space Administration under the Exoplanet Exploration Program.

This work has made use of data from the European Space Agency (ESA) mission {\it Gaia} (\url{https://www.cosmos.esa.int/gaia}), processed by the {\it Gaia} Data Processing and Analysis Consortium (DPAC, \url{https://www.cosmos.esa.int/web/gaia/dpac/consortium}). Funding for the DPAC has been provided by national institutions, in particular the institutions participating in the {\it Gaia} Multilateral Agreement.

This article is partly based on observations made with the MuSCAT2 instrument, developed by ABC, at Telescopio Carlos S\'{a}nchez operated on the island of Tenerife by the IAC in the Spanish Observatorio del Teide.
This work makes use of observations from the LCOGT network.
This work makes use of observations acquired with the T150 telescope at Sierra Nevada Observatory, operated by the Instituto de Astrof\'{i}sica de Andaluc\'{i}a (IAA-CSIC).
Some of the Observations in the paper made use of the High-Resolution Imaging instrument `Alopeke at Gemini-North. `Alopeke was funded by the NASA Exoplanet Exploration Program and built at the NASA Ames Research Center by Steve B. Howell, Nic Scott, Elliott P. Horch, and Emmett Quigley. 
IRD is operated by the Astrobiology Center of the National Institutes of Natural Sciences. 
The research leading to these results has received funding from  the ARC 
grant for Concerted Research Actions, financed by the Wallonia-Brussels Federation. TRAPPIST is funded by the Belgian Fund for Scientific Research (Fond National de la Recherche Scientifique, FNRS) under the grant FRFC 2.5.594.09.F, with the participation of the Swiss National Science Fundation (SNF). TRAPPIST-North is a project funded by the University of Liege (Belgium), in collaboration with Cadi Ayyad University of Marrakech (Morocco) MG and EJ are F.R.S.-FNRS Senior Research Associate. 
The authors acknowledge funding from the Spanish Ministry of Economics and Competitiveness through projects PGC2018-098153-B-C31 and AYA2015- 69350-C3-2-P. This work is partly supported by JSPS KAKENHI Grant Numbers JP18H01265 and JP18H05439, and JST PRESTO Grant Number JPMJPR1775. V.M.P. acknowledges support from NASA Grant NNX17AG24G. T.H. acknowledges support from the European Research Council under the Horizon 2020 Framework Program via the ERC Advanced Grant Origins 83 24 28. This research has been partially funded by Project No. MDM-2017-0737 Unidad de Excelencia "Mar\'{i}a de Maeztu" - Centro de Astrobiolog\'{i}a (INTA-CSIC). This research  acknowledges financial support from the State Agency for Research of the Spanish MCIU through the "Center of Excellence Severo Ochoa" award to the Instituto de Astrof\'{i}sica de Andaluc\'{i}a (SEV-2017-0709) and project AYA2016-79425.
\end{acknowledgements}


\appendix
\section{Long tables}
\label{apx-long_tables}
\begin{table}
\caption{Radial velocities of \host.}
\label{table-rvs}
\begin{tabular}{lrr}
\hline
\hline
BJD$_\mathrm{TDB}$-2\,450\,000  & RV (\mps)       & $\sigma_\mathrm{RV}$ (\mps)\\
\hline
\noalign{\smallskip}
\multicolumn{3}{c}{3.5\,m Calar Alto/CARMENES-VIS}\\
8844.63067 &        3.676 &        5.861\\
8844.75146 &       -7.469 &        3.631\\
8846.60074 &       -1.753 &        1.615\\
8846.71780 &       -2.432 &        1.258\\
8848.62456 &        5.964 &        1.383\\
8848.73609 &        8.577 &        1.335\\
8849.59266 &        5.041 &        1.453\\
8849.73669 &        1.953 &        1.648\\
8850.61843 &        1.189 &        1.191\\
8850.72764 &       -1.199 &        1.331\\
8851.63152 &        0.324 &        1.133\\
8851.74504 &        3.429 &        1.170\\
8852.65899 &        5.364 &        1.325\\
8852.75676 &       -0.902 &        1.097\\
8854.61164 &       -8.242 &        1.798\\
8855.60409 &        0.717 &        1.543\\
8855.74490 &       -2.347 &        1.574\\
8856.58282 &       -2.463 &        1.401\\
8856.72077 &       -8.844 &        1.582\\
8857.57560 &       -7.257 &        1.523\\
8858.63243 &        1.425 &        2.106\\
8860.55597 &        6.827 &        2.478\\
8860.67971 &        1.469 &        2.199\\
8864.56029 &       -4.091 &        1.859\\
8865.55825 &        6.382 &        2.696\\
8865.67852 &        3.557 &        1.896\\
8866.61374 &        2.104 &        2.638\\
8877.52010 &        0.316 &        1.954\\
8877.68512 &       -1.075 &        2.024\\
8878.57708 &       -1.038 &        1.918\\
8880.63111 &       -5.071 &        1.894\\
8880.71831 &       -2.735 &        2.399\\
8881.50534 &       -2.453 &        1.370\\
8881.65959 &       -0.782 &        1.341\\
8882.48468 &       -1.096 &        1.643\\
8882.67088 &        1.903 &        1.530\\
8883.52731 &        1.816 &        1.454\\
8883.68305 &       -2.280 &        1.520\\
8884.52841 &       -0.715 &        1.440\\
8884.67457 &        2.912 &        1.432\\
8885.51388 &        6.990 &        1.509\\
8885.63103 &        6.912 &        1.494\\
8887.54307 &       -3.888 &        3.347\\
8889.60111 &        0.741 &        1.967\\
8890.55909 &       -0.965 &        1.680\\
8891.50814 &       -6.148 &        1.589\\
8893.55000 &       -1.000 &        1.539\\
8894.51095 &       -8.253 &        1.487\\
8895.52663 &       -0.796 &        1.523\\
8896.50587 &        0.123 &        1.674\\
8897.47151 &       -1.596 &        1.627\\
8898.57582 &       -5.761 &        2.995\\
\hline
\noalign{\smallskip}
\multicolumn{3}{c}{Subaru/IRD}\\
8828.06968 &        1.950 &        2.460\\
8828.07559 &        1.240 &        2.460\\
8828.08218 &        4.990 &        2.460\\
8828.08811 &       -2.390 &        2.410\\
\hline
\noalign{\smallskip}
\multicolumn{3}{c}{IRTF/iSHELL}\\
8874.08622 &       -9.389 &       11.841\\
8875.02099 &       -4.854 &        9.743\\
8894.99312 &        4.020 &       10.855\\
8898.96755 &      -14.658 &        5.359\\
8899.95713 &      -17.587 &        8.026\\
8900.95157 &       -0.664 &        7.871\\
8907.99311 &       11.516 &       10.568\\
8911.02803 &        5.116 &        7.481\\
\hline
\end{tabular}
\end{table}

\section{Joint modelling}
\label{sec:joint_modelling_posteriors}

\subsection{Model parameter priors}
\begin{table*}
\centering
\caption{Parameter priors for the combined modelling of transit light curves and RV observations.} 
\label{tab:priors}
\begin{tabular}{lccl}
\hline
\hline
\noalign{\smallskip}
Parameter name & Prior & Unit & Description \\
\noalign{\smallskip}
\hline
\noalign{\smallskip}
Parameters for planet b \\[0.1 cm] 
\noalign{\smallskip}
~~~$P_{b}$  & $\mathcal{N}(0.768418, 0.0002)$ & d & Orbital period\\[0.1 cm]
~~~$t_{0,b}$  & $\mathcal{N}(2458543.9115, 0.0036)$ & d & Time of transit centre\\[0.1 cm]
~~~$k^2_b$  & $\mathcal{U}(0.0, 0.01)$ & ... & True area ratio\\[0.1 cm]
~~~$b_b$  & $\mathcal{U}(0.0, 1.0)$ & ... & Impact parameter\\[0.1 cm]
~~~$K_{b}$  & $\mathcal{U}(0.0, 36.0)$ & m\,s$^{-1}$ & Radial velocity semi-amplitude\\[0.1 cm]
~~~$S_{1,b} = \sqrt{e_b}\sin \omega_b$  &$\mathcal{U}(-0.5, 0.5)$   & ... & Parametrisation for $e$ and $\omega$ \\[0.1 cm]
~~~$S_{2,b} = \sqrt{e_b}\cos \omega_b$  & $\mathcal{U}(-0.5, 0.5)$  & ... & Parametrisation for $e$ and $\omega$ \\[0.1 cm]
\noalign{\smallskip}
Parameters for planet c \\[0.1 cm] 
\noalign{\smallskip}
~~~$P_{c}$  & $\mathcal{N}(12.254218, 0.006^2)$ & d & Orbital period\\[0.1 cm]
~~~$t_{0,c}$  & $\mathcal{N}(2458546.8484, 0.0042^2)$ & d & Time of transit centre\\[0.1 cm]
~~~$k^2_c$  & $\mathcal{U}(0.0, 0.01)$ & ... & True area ratio\\[0.1 cm]
~~~$b_c$  & $\mathcal{U}(0.0,1.0)$ & ... & Impact parameter\\[0.1 cm]
~~~$K_{c}$  & $\mathcal{U}(0.0,36.0)$ & m\,s$^{-1}$ & Radial velocity semi-amplitude\\[0.1 cm]
~~~$S_{1,c} = \sqrt{e_c}\sin \omega_c$  & $\mathcal{U}(-0.5, 0.5)$  & ... & Parametrisation for $e$ and $\omega$ \\[0.1 cm]
~~~$S_{2,c} = \sqrt{e_c}\cos \omega_c$  & $\mathcal{U}(-0.5, 0.5)$  & ... & Parametrisation for $e$ and $\omega$ \\[0.1 cm]
\noalign{\smallskip}
Flux contamination parameters\\[0.1 cm]
\noalign{\smallskip}
~~~$c_{{TESS}}$  & $\mathcal{U}(0.0, 0.99)$  & ... & {\em TESS} contamination\\[0.1 cm]
~~~$c_{\textnormal{gb}}$  & $\mathcal{U}(0.0, 0.99)$  & ... & Ground-based contamination in $r'$\\[0.1 cm]
~~~$T_{\mathrm{eff},h}$  & $\mathcal{N}(3360.0, 51)$  & K & Host star $T_\mathrm{eff}$\\[0.1 cm]
~~~$T_{\mathrm{eff},c}$  & $\mathcal{U}(2500, 12000)$  & K & Contaminant $T_\mathrm{eff}$\\[0.1 cm]
\noalign{\medskip}
\noalign{\smallskip}
Stellar parameters\\[0.1 cm]
\noalign{\smallskip}
~~~$\rho_{*}$  & $\mathcal{N}(9.6, 1.0)$ & g\,cm$^{-3}$ & Stellar density\\[0.1 cm]
~~~$q_{1,f}$  & \texttt{LDTk} & ... & Limb-darkening q$_1$ for passband $f$\\[0.1 cm]
~~~$q_{2,f}$  & \texttt{LDTk} & ... & Limb-darkening q$_2$ for passband $f$\\[0.1 cm]
\noalign{\medskip}
Additional parameters\\[0.1 cm]
\noalign{\smallskip}
~~~$\log e_i$  & $\mathcal{U}(-4, 0)$ & - & $\log10$ average white noise for light curve $i$\\[0.1 cm]
~~~$z_i$  & $\mathcal{N}$ & ... & Baseline constant (intercept) for light curve $i$\\[0.1 cm]
~~~$c_{i,j}$  & $\mathcal{N}$ & ... & Baseline coefficient for light curve $i$ covariate $j$\\[0.1 cm]
\noalign{\smallskip}
RV instrumental parameters\\[0.1 cm]
\noalign{\smallskip}
~~~$\gamma_{\textnormal{CARMENES}}$  & $\mathcal{U}(-100,100)$ & m\,s$^{-1}$ & Systemic velocity for CARMENES\\
~~~$\gamma_{\textnormal{IRD}}$  & $\mathcal{U}(-100,100)$ & m\,s$^{-1}$ & Systemic velocity for Subaru/IRD\\
~~~$\gamma_{\textnormal{iSHELL}}$  & $\mathcal{U}(-100,100)$ & m\,s$^{-1}$ & Systemic velocity for IRTF/iSHELL\\[0.1 cm]
\noalign{\smallskip}
\hline
\end{tabular}
\end{table*}

\subsection{Posterior distribution functions}
\begin{figure*}
\centering
\includegraphics[width=\linewidth]{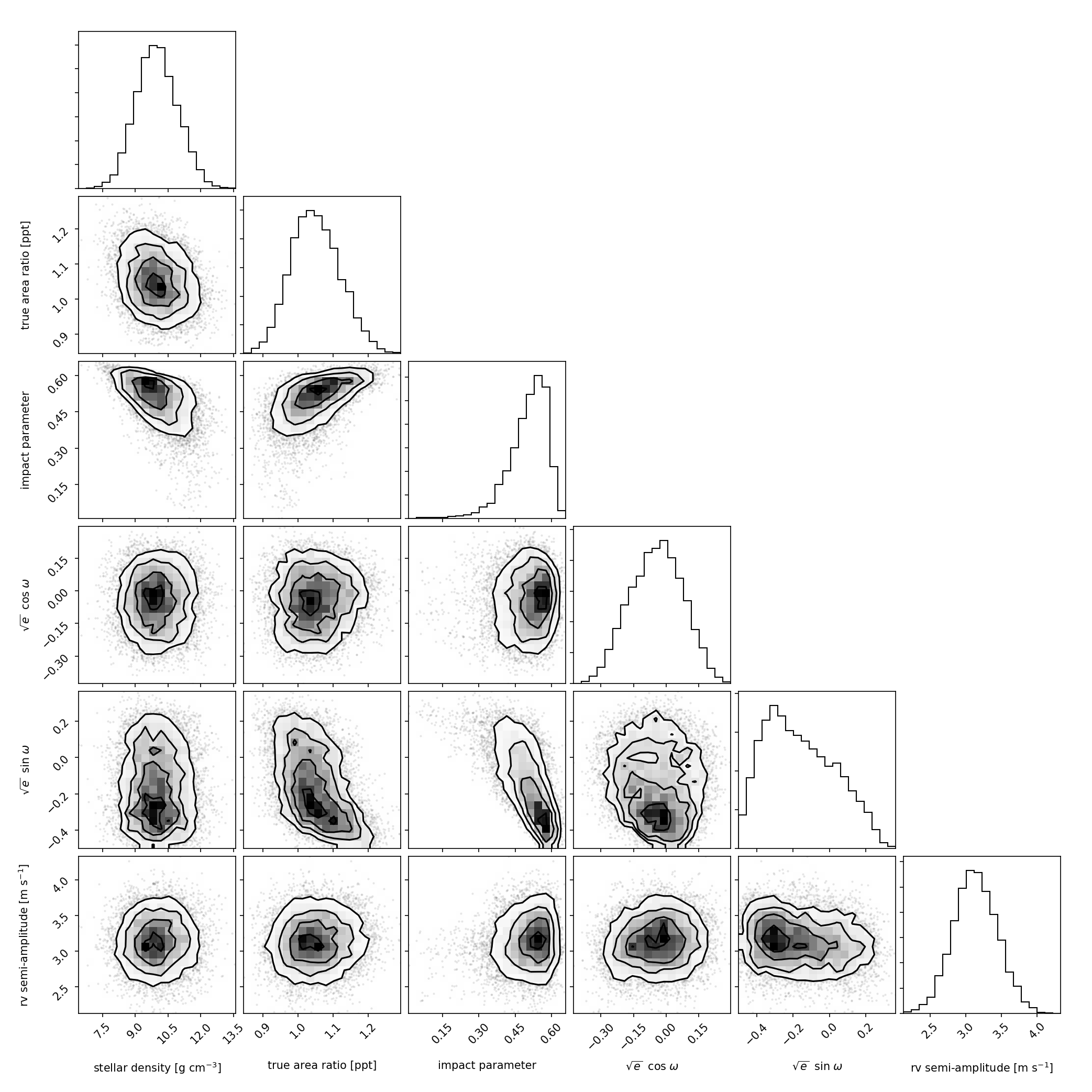}
\caption{Marginal and joint parameter posterior densities for the sampling parameters 
describing \host{} b.}
\label{fig:corner_01}
\end{figure*}

\begin{figure*}
\centering
\includegraphics[width=\linewidth]{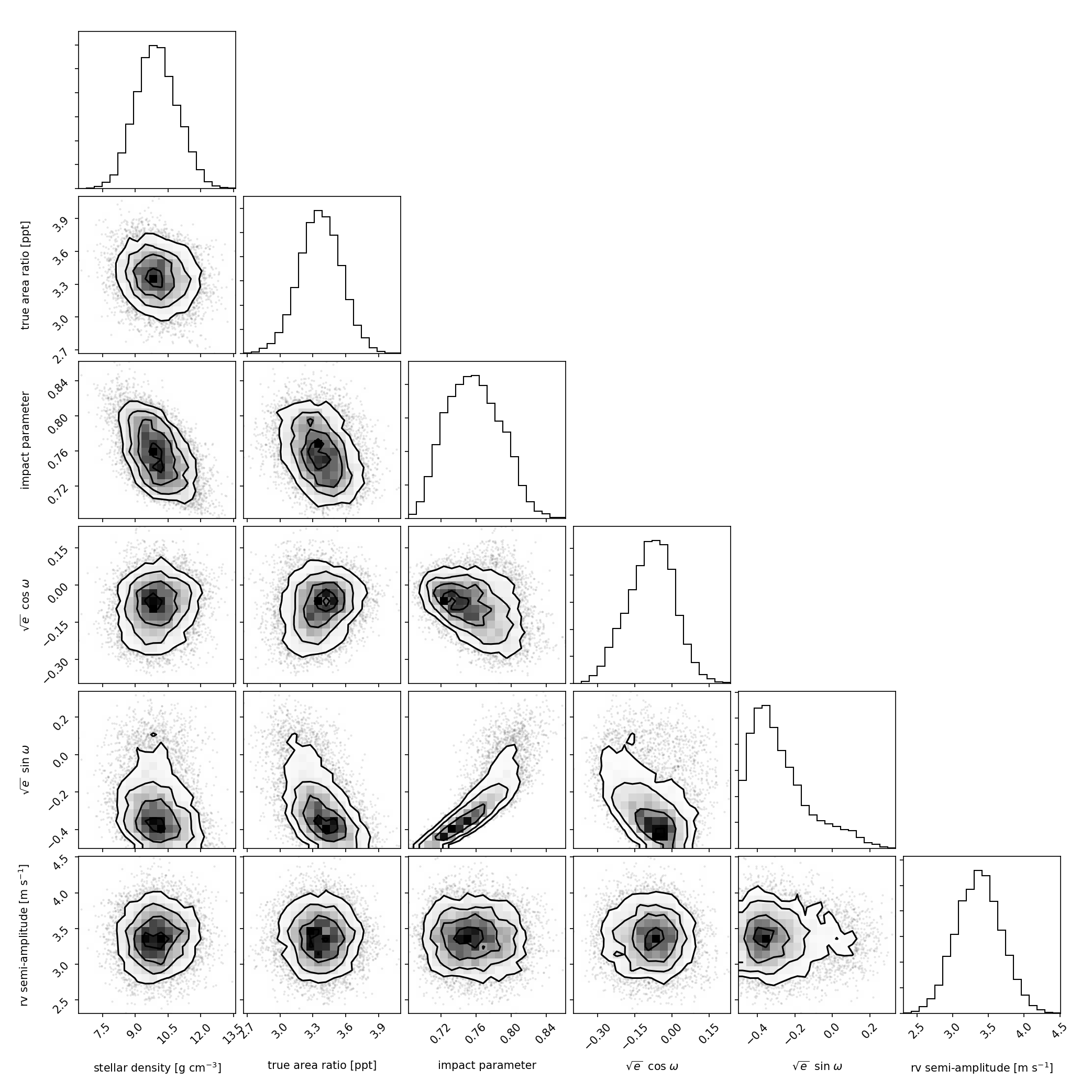}
\caption{Marginal and joint parameter posterior densities for the sampling parameters 
describing \host{} c.}
\label{fig:corner_02}
\end{figure*}

\end{document}